\documentclass[%
 reprint,
 amsmath,amssymb,
 aps,
]{revtex4-2}

\usepackage{graphicx}% Include figure files
\usepackage{dcolumn}% Align table columns on decimal point
\usepackage{bm}% bold math
\usepackage{amsmath,latexsym,amssymb, amsfonts}
\usepackage{slashed}

\def\vec#1{\mathchoice
 {\mbox{\boldmath $\displaystyle#1$}}
 {\mbox{\boldmath $\textstyle#1$}}
 {\mbox{\boldmath $\scriptstyle#1$}}
 {\mbox{\boldmath $\scriptstyle#1$}}}

\newcommand{\Slash}[1]{{\ooalign{\hfil/\hfil\crcr$#1$}}}

\usepackage{ulem}
\usepackage{color}
\usepackage{cancel}

\begin{document}

\preprint{APS/123-QED}

\title{Quantum molecular dynamics model based on relativistic mean field theory for light nucleus fragmentation in hadron therapy}% Force line breaks with \\

%\affiliation{%

\author{
Akihiro Haga$^{1,*}$
Yoshi-hide Sato$^{1}$, 
Hana Fujiwara$^{1}$, 
        Dousatsu Sakata$^{2, 3, 4}$, \\
        David Bolst$^{3}$, 
        Edward C. Simpson$^{5}$, 
        and Susanna Guatelli$^{3}$
        }

\address{1 Graduate School of Biomedical Sciences, Tokushima University, Tokushima 770-8503, Japan}
%\address{2 National Institute of Advanced Industrial Science and Technology, Ibaraki 305-8560 , Japan} % I can not use this affiliation for now
\address{2 Division of Health Science, University of Osaka, Suita 565-0871, Japan}
\address{3 Centre For Medical and Radiation Physics, University of Wollongong, Wollongong NSW 2522, Australia}
\address{4 School of Physics, University of Bristol, Bristol, BS8 1TL, United Kingdom}
\address{5 Department of Nuclear Physics and Accelerator Applications, Research School of Physics, The Australian National University, Canberra ACT 2601, Australia}
%\address{6 Australian Nuclear Science and Technology Organisation (ANSTO), NSW, Australia}
%\address{7 Brain and Mind Centre, University of Sydney, Sydney, NSW, Australia}
\email{haga@tokushima-u.ac.jp}


%\affiliation{
%\affiliation{
%\affiliation{%


\date{\today}% It is always \today, today,

\begin{abstract}
This study evaluates the accuracy of nuclear fragmentation simulations using a quantum molecular dynamics (QMD) model based on relativistic mean field (RMF) theory for an energy range of 50–400 MeV/u, relevant to hadron therapy.
A total of 10 parameter sets within the RMF framework are assessed based on their ability to reproduce ground-state properties such as the mean squared radius and binding energy, as obtained in QMD simulations. Among these, the NS2 parameter set is identified as the most suitable for describing stable nuclei over a wide mass range, with the use of an adaptive Gaussian wave packet width. 
%Using NS2, additional tuning of parameters, including the maximum time for evolution, maximum impact parameter, and clustering distance, is performed to optimize the model.
Fragmentation cross sections of carbon ion projectiles on light nuclei targets (H, C, O, Al, Ti, and Cu) are simulated at incident energies of 50, 95, 290, and 400 MeV/u and compared with experimental data. The results indicate that the RQMD.RMF model provides superior reproductions for fragmentation at lower energies (50 and 95 MeV/u) compared to the Light Ion QMD (LIQMD) model implemented in Geant4 version 11.2. At higher energies (290 and 400 MeV/u), the RQMD.RMF model performs comparably to the LIQMD.
This study demonstrates that the RQMD.RMF model provides a reliable framework for analyzing nuclear fragmentation and holds potential for applications in the planning and quality assurance of hadron therapy.
%An article usually includes an abstract, a concise summary of the work
%covered at length in the main body of the article. 
%\begin{description}
%use the optional argument of the \verb+\item+ command to give %the category of each item. 
%\end{description}
\end{abstract}

\maketitle

%\tableofcontents

\section{\label{sec:intro}Introduction}
The relativistic version of the quantum molecular dynamics (QMD) model, known as RQMD, has been developed with a Lorentz scalar treatment\cite{Sorge1989, Maruyama1991, Maruyama1996}. This advancement demonstrated the importance of including relativistic effects for accurately simulating nuclear fragmentation\cite{Ogawa2018} and transverse flow\cite{Maruyama1995, Lehmann1995} even in the intermediate energy range ($\sim$1 GeV/u). While these studies mainly utilized the Skyrme interaction, which falls under the category of non-relativistic nuclear models, recent progress has shifted the focus toward relativistic approaches\cite{Nara2019,Nara2020,Wei2024}.

A QMD model based on the relativistic mean field (RMF) theory, termed RQMD.RMF, has been developed by Nara et al.\cite{Nara2019,Nara2020}. This model employs a fully covariant framework with self-consistent scalar and vector potentials via meson interactions. Initial investigations with RQMD.RMF have explored the sensitivity of collective flow excitation functions in heavy nucleus collisions, such as Au+Au, in high-energy regions ($2.5 \le \sqrt{s_{NN}} \le 20$ GeV). However, there has been little to no investigation into light nucleus collisions, such as C+C, in the energy range (50–400 MeV/u), which is particularly relevant for applications in hadron therapy\cite{Schardt96, Matsufuji03, Haettner06, Bolst17}.

Monte Carlo based simulations have become increasingly relevant in the medical field, especially, of hadron therapy\cite{Arce2021, Arce2025}, where the precise modeling of nuclear reactions and fragmentation processes is critical for treatment accuracy and safety. In hadron therapy, light ions such as carbon or oxygen are used to deliver localized radiation to tumors while minimizing damage to healthy tissue surrounding the tumors. However, secondary fragments generated during the irradiation can travel beyond the intended target region, potentially affecting healthy tissues. The ability of QMD models to simulate these fragmentation processes with high accuracy makes them an essential tool for evaluating the therapeutic efficacy and optimizing treatment planning in hadron therapy. Moreover, an accurate modeling of nuclear fragmentation is also very important to develop novel imaging applications used for hadron therapy, such as positron emission tomography\cite{Rahmim13, Bertolli16, Lestand2017, Hofmann2019, Chacon19, Horst2019, Pratt2023, Chacon2024} and/or prompt gamma-ray measurement\cite{Dedes2014, Pinto2015, Mattei2015, Missaglia2023, Pratt2023} for range monitoring. Therefore, the quantitative assessment of nuclear fragmentation using the RQMD.RMF model is of great interest, not only in the context of high-energy physics but also for advancing applications in medical physics, such as improving the precision and safety of hadron therapy.

The Walecka-type RMF model\cite{Serot1992}, which incorporates interactions among mesons, has successfully provided an appropriate description of the incompressibility of nuclear matter\cite{Horowitz1987} and surface properties in finite nuclei\cite{Ring1996}. Over time, the RMF model has been extended and widely applied to the study of both normal nuclei and hypernuclei, leading to the proposal of many effective interaction parameter sets\cite{Reinhard1986, Patra1993, Lalazissis1997, Lalazissis2009, Sharma1993, Sugahara1994, long2004}. Notably, RMF models have demonstrated significant success in reproducing ground-state properties across a broad mass range, including light nuclei\cite{Rong2023}. 

In this study, we aim to expand the application of the RQMD.RMF model by investigating the fragmentation cross-sections relevant to carbon ion therapy, specifically at energies of 50, 95, 290, and 400 MeV/u. These energies are studied using various targets, including H, C, O, Al, and Ti for 50 and 95 MeV/u, and H, C, Al, and Cu for 290 and 400 MeV/u. In the preliminary stage of this research, 10 RMF parameter sets proposed thus far are assessed in terms of their ability to reproduce ground-state properties using QMD simulations. This evaluation enables the selection of a single parameter set for application to fragmentation cross-section calculations. By bridging the gap in understanding light nucleus collisions in the range of 50-400 MeV/u, this study seeks to provide valuable insights into both fundamental nuclear physics and practical applications in hadron therapy.

\section{\label{sec:RMFQMD}RQMD.RMF model}
Here, we describe the theoretical framework of QMD based on the RMF model. We begin with a brief review of the relativistic QMD model formulated using the constraint Hamiltonian approach\cite{Komar1978-01, Komar1978-02, Komar1978-03, Marty2013, Nara2019}. Then, we derive the formulations of the relativistic nuclear model, specifically the RQMD.RMF model.
Details of the nuclear ground state properties and the nuclear fragmentation cross sections used for model validation are also provided.

\subsection{\label{sec:formalism}Relativistic QMD}
The relativistic version of the QMD model, RQMD, is formulated based on the constraint Hamiltonian dynamics\cite{Komar1978-01, Komar1978-02, Komar1978-03}, which reduce the number of degrees of freedom in the relativistic phase space (8$N$ $\rightarrow$ 6$N$)\cite{Maruyama1996, Mancusi09, Nara2019}. World lines are given by $(\vec{q}_i(\tau), \vec{p}_i(\tau))$, meaning that position and momentum of $i$-th particle as a function of the time $\tau$\cite{Marty2013}.
The constraints $\phi_k, k \in 1 \sim 2N$ (including time constraints) are assumed to be invariant,
\begin{align}
\frac{d\phi_i}{d\tau} = \frac{\partial \phi_i}{\partial \tau} + \sum_{k=1}^{2N} C_{ik}^{-1} \lambda_k  = 0,
\label{eq1}
\end{align}
on the physical space, where $C_{ik}^{-1} \equiv \{\phi_i, \phi_k\}$ is the Poisson brackets for four-vectors:
\begin{align}
\{\phi_i, \phi_k\} = \sum_{k'=1}^{N}
\left(
\frac{\partial\phi_i}{\partial q_{k'}^\mu}\frac{\partial\phi_k}{\partial p_{k'\mu}}
-\frac{\partial\phi_i}{\partial p_{k'}^\mu}\frac{\partial\phi_k}{\partial q_{k'\mu}}
\right)
.
\end{align}
Now, we constrain as,
\begin{align}
\phi_k = 
\left\{
\begin{array}{r}
 K_k (q^\mu, p^\mu) = 0, \hspace{2.0cm}(k \in (1 \sim N)) \\ 
\chi_{k} (q^\mu, p^\mu) = 0, \hspace{0.5cm}(k \in (N+1 \sim 2N-1))  \\
\chi_{2N} (q^\mu, p^\mu, \tau) = 0, \hspace{2.2cm}(k=2N)
\end{array} 
\right.
\end{align}
where only $\phi_{2N}$ depends on $\tau$. 
We use the on-mass shell condition, $K_k = p_k^{*2} - m_k^{*2}$, for $k \in (1 \sim N)$ and time constraints,
\begin{align}
\chi_{i+N} &= U \cdot (q_{i} - q_N), \hspace{2mm}(i \in 1 \sim N-1) \\ 
\chi_{2N} &= U \cdot q_{N} - \tau, 
\end{align}
for others. Here $U = (1,\vec{0})$ is applied \cite{Mancusi09, Nara2019, Nara2020}. 
Assuming $\{K_{i}, K_{j}\} = 0$,
%the equation $\sum_k^{2N} C_{i,k}^{-1}\lambda_k = -\partial \phi_i/\partial \tau$ 
Eq.~(\ref{eq1})
gives,
\begin{align}
\sum_{k=N+1}^{2N} C_{i,k}^{-1} \lambda_k &= 0,
\hspace{2mm}(i \in 1 \sim N)
\end{align}
then, we have $\lambda_k = 0$ for $k\in (N+1 \sim 2N$).
This leads,
\begin{align}
\sum_{k=1}^{N} C_{i+N,k}^{-1} \lambda_k &= 2 p_i^{0*} \lambda_i - 2 p_N^{0*} \lambda_N = 0,\\
\sum_{k=1}^{N} C_{2N,k}^{-1} \lambda_k &= 2 p_N^{0*} \lambda_N = 1,
\end{align}
for $i\ne N$ and $i=N$, respectively. Thus, we apply 
$\lambda_k = 1/(2p_k^{*0})$ and $\lambda_{k+N} = 0$ for $k = 1,...,N$ to the equation of motion with the present condition,

%Assuming $\{K_i, K_j\} = 0$\cite{Sorge1989}, we have,
%\begin{align}
%\lambda_k = -C_{k,2N}\frac{\partial \phi_{2N}}{\partial \tau},
%\end{align}
%and $\lambda_k = 0$ for $k \in (N+1 \sim 2N)$.
%We define the constraint Hamiltonian as $\sum_{k=1}^{2N} \lambda_k \phi_k$, then, the equations of motion are given by,
\begin{align}
\frac{dq_i^\mu}{d\tau} = & \sum_{k=1}^N 
\frac{1}{2p^{0*}_k}
%\lambda_k 
\frac{\partial K_k}{\partial p_{i\mu}}, 
\label{eomRQMD1} \\
\frac{dp_i^\mu}{d\tau} = & -\sum_{k=1}^N 
\frac{1}{2p^{0*}_k}
%\lambda_k 
\frac{\partial K_k}{\partial q_{i\mu}}.
\label{eomRQMD2}
\end{align}
%where the Lagrange multipliers $\lambda_k$ has an analytical solution\cite{Marty2013}, 
%\begin{align}
%\lambda_k = (2p_k^{*\mu} U_\mu)^{-1} = \frac{1}{2p^{0*}_k},
%\label{lambdak}
%\end{align}
%with $U_\mu \equiv (1,\vec{0})$, where
%$p_k^{*\mu}$ includes the vector potential $\Sigma_k^\mu$ as 
%$p_k^\mu-\Sigma_k^\mu$.
%The constraint $K_k$ taken by on-mass shell condition:
%\begin{align}
%K_k = p_k^{*2}-m_k^{*2}, \hspace{3mm} k \in (1 \sim N),
%\label{Kk}
%\end{align}
%{\color{red}includes
%the vector potential $\Sigma_{v,k}^\mu$ as 
%$p_k^\mu-\Sigma_{v,k}^\mu$
%and the scalar potential
%$\Sigma_{s,k}$
%as} $m_k^* = m_k-\Sigma_{s,k}$. %includes the scalar potential $\Sigma_{s,k}$.

It has been demonstrated in Ref.\cite{Maruyama1996} that this RQMD approach, which uses the common time coordinate for all particles, can be used up to 6 GeV/u. In the RQMD using a non-relativistic Skyrme type interaction, the constraint and energy are considered as,
\begin{align}
K_i \equiv& 
\hspace{1mm}
p_i^2 - m_i^2 - 2m_i V_{i}, 
\\
p^0_i \equiv&
\sqrt{\vec{p}_i^2 + m_i^2 + 2m_i V_{i}},
\label{RQMDE}
\end{align} 
which is regarded as the Lorentz scalar treatment of the Skyrme potential $V_{i}$. In this study, we update the formula to include both scalar and vector potentials as,
\begin{align}
K_i \equiv& 
\hspace{1mm}
p_i^{*2} - m_i^{*2}, 
\label{Kk}\\
p^{0}_i \equiv & \sqrt{(\vec{p}_i-\vec{\Sigma}_{v,i})^2 + (m_i-\Sigma_{s,i})^2 } + V_i, \\
 & V_i = 
\frac{1}{2}f_i\Sigma_{s,i}
+ \frac{1}{2}\Sigma_{v,i}^0
+ \frac{1}{2}
%\frac{\vec{p}_i}{p_i^0}
\vec{\beta}_i
\cdot \vec{\Sigma}_{v,i}
+ ({\rm SE})_i,
\end{align} 
where 
$f_i = m_i/\sqrt{\vec{p}_i^2 + m_i^2}, \vec{\beta}_{i} = \vec{p}_{i}/\sqrt{\vec{p}_i^2 + m_i^2}$,
and
SE is the meson self-energy.
We define the four momentum in nuclear medium with,
\begin{align}
p_i^{*\mu} = p_i^{*\mu}-\tilde{\Sigma}_{v,i}^\mu, 
\hspace{2mm}
\tilde{\Sigma}_{v,i}^\mu =
(V_i, \vec{\Sigma}_{v,i}).
\end{align}
As shown below, the energy $p^{0}_i$ is equivalent to the energy component of energy--momentum tensor of the RMF model.
%{\color{red}
The concrete formula of $\Sigma_{s,i}$, $\Sigma_{v,i}^\mu$, and (SE)$_i$ are also given in the next subsection.
%}
The equations of motion in RQMD is obtained by substituting Eq.~(\ref{Kk}) into Eqs.(\ref{eomRQMD1})
and (\ref{eomRQMD2}), 
\onecolumngrid
%{\color{red}
\begin{align}
\frac{dq_i^\mu}{d\tau} &=  {\beta_i^*}^\mu +
\sum_{k=1}^N 
\left[
\left(
f_k^* - \frac{f_k}{2}
\right)
\frac{\partial\Sigma_{s,k}}{\partial p_{i\mu}} 
-
\left(
\beta_{k\nu}^* - \frac{\beta_{k\nu}}{2}
\right)
\frac{\partial{\Sigma}_{v,k}^\nu}{\partial p_{i\mu}}
 -
\frac{\partial({\rm SE})_k}{\partial p_{i\mu}}
-
\frac{1}{2}\frac{\partial f_k}{\partial p_{i\mu}}\Sigma_{s,k}
+
\frac{1}{2}\frac{\partial \beta_{k\nu}}{\partial p_{i\mu}}\Sigma_{v,k}^\nu
\right],
\label{EoM1}\\
\frac{dp_i^\mu}{d\tau} &=  
- \sum_{k=1}^N 
\left[
\left(
f_k^* - \frac{f_k}{2}
\right)
\frac{\partial\Sigma_{s,k}}{\partial q_{i\mu}}
-
\left(
\beta_{k\nu}^* - \frac{\beta_{k\nu}}{2}
\right)
\frac{\partial{\Sigma}_{v,k}^\nu}{\partial q_{i\mu}}
-
\frac{\partial({\rm SE})_k}{\partial q_{i\mu}}
\right] \label{EoM2},
\end{align}
\twocolumngrid
\noindent
where $f_k^* = m_k^*/p_k^{*0}$
and $\beta_{k\nu}^* = p_{k\nu}^*/p_k^{*0}$.
The equation of motion simulation
using above formula requires the derivatives of $\Sigma_{s,k}, \Sigma_{v,k}^\nu, ({\rm SE})_k, f_k$, and $\beta_{k\nu}$ in $\vec{p}_i$ and $\vec{q}_i$, and these are given in Appendix.
%}

\subsection{RMF model in QMD}
We employ the Walecka-type RMF model\cite{Horowitz1987,Serot1992} including the mean field of $\sigma$, $\omega$, $\vec{\rho}$ mesons and photon, where the $\sigma$ 
%{\color{red} \sout{and $\omega$ mesons are} 
meson is 
%} 
allowed to have self-energy (non-linear) interaction terms,
as the Lagrangian:
\begin{align}
{\cal L} =& \bar{\psi}[i \Slash{\partial} - m_N + g_\sigma \sigma - g_\omega \Slash{\omega} - g_\rho \vec{\tau}\cdot \Slash{\vec{\rho}} - eQ \Slash{A}]\psi \nonumber\\
& + \frac{1}{2} \left( \partial_\mu \sigma \right)^2 - \frac{1}{2} m_\sigma^2 \sigma^2 
- \frac{1}{3}g_2 \sigma^3 - \frac{1}{4}g_3 \sigma^4 \nonumber\\
& - \frac{1}{4} \left( \Omega_{\mu\nu}\right)^2 + \frac{1}{2} m_\omega^2 \omega_\mu \omega^\mu  
%{\color{red}
%\cancel{+ \frac{1}{4} c_3 (\omega_\mu \omega^\mu)^2 }}
\nonumber\\
& - \frac{1}{4} \left( \vec{R}_{\mu\nu}\right)^2 + \frac{1}{2} m_\rho^2 \vec{\rho}_\mu\cdot \vec{\rho}^\mu  \nonumber\\
& - \frac{1}{4} \left( F_{\mu\nu}\right)^2,
\end{align}
where,
\begin{align}
\Omega_{\mu\nu} =& \partial_\mu \omega_\nu - \partial_\nu \omega_\mu, \\
%\vec{R}_{\mu\nu} =& \partial_\mu \vec{\rho}_\nu - \partial_\nu \vec{\rho}_\mu -g_\rho \epsilon^{abc}\rho_\mu^b\rho_\nu^c, \\
R_{\mu\nu}^a =& \partial_\mu^a \rho_\nu^a - \partial_\nu \rho_\mu -g_\rho \epsilon^{abc}\rho_\mu^b\rho_\nu^c, \\
F_{\mu\nu} =& \partial_\mu A_\nu - \partial_\nu A_\mu,
\end{align}
and the $\vec{\tau}$ is the isospin matrix and $Q=(1+\tau_3)/2$ is the charge operator. The Hamiltonian in the system can be derived from the $\{00\}$ component of the energy-momentum tensor ($T_{\mu\nu} = - g_{\mu\nu} {\cal L} + (\partial {\cal L}/\partial(\partial_\mu \phi)) (\partial_\nu \phi)$),
\begin{align}
\int d\vec{r} T_{00} = 
\int d\vec{r} \left( -{\cal L} + \frac{\partial {\cal L}}{\partial(\partial_0 \phi)} (\partial_0 \phi) \right)
\equiv H,
\end{align}
%{\color{red}
where the Hamiltonian is then:
%}
\begin{align}
H = & \sum_i \left(
\sqrt{{\vec{p}_i^*}^2 + {m_i^*}^2} + \frac{1}{2}\Sigma_{v,i}^0 + \frac{1}{2}\vec{\Sigma}_{v,i}\cdot \vec{\beta}_i + \frac{1}{2}\Sigma_{s,i} f_i 
\right)\nonumber \\
&+ \int d\vec{r} \left(
- \frac{1}{6}g_2 \sigma^3 - \frac{1}{4}g_3 \sigma^4 
%{\color{red}\cancel{
%+ \frac{1}{4}c_3 (\omega_\mu\omega^\mu)^2
%}}
\right),
\end{align}
where the last term shows the self energy of $\sigma$ %{\color{red}\sout{ and $\omega$}} 
meson field arising from the non-linear terms. 
%{\color{red}\sout{, and, }
%\begin{align}
%\cancel{
%\vec{p}_i^* = \hspace{1mm} \vec{p}_i - \vec{\Sigma}_{v,i},
%}
%\nonumber
%\\
%\cancel{
%m_i^* = \hspace{1mm} m_N - \Sigma_{s,i}, }
%\nonumber
%\\
%\cancel{
%\vec{\beta}_i \hspace{5mm}= \hspace{5mm} \frac{\vec{p}_i}{p^0_i}\hspace{5mm}, 
%}
%\nonumber
%\\
%\cancel{
%f_i \hspace{5mm}= \hspace{5mm} \frac{m_N}{p^0_i}\hspace{5mm},
%}
%\nonumber
%\end{align}
%\sout{and $p^0_i = \sqrt{{\vec{p}_i}^2 + {m_N}^2}$. }}
$\Sigma_{s,i}$ is the scalar potential arising from the scalar meson to $i$-th particle, while $\Sigma_{v,i}^0$ and $\vec{\Sigma}_{v,i}$ are the time and the space components of the vector potential to $i$-th particle, respectively. 
The mean field potentials can be obtained from the Euler-Lagrange equation ($\partial {\cal L}/\partial \phi - \partial_\mu \partial {\cal L}/\partial (\partial_\mu \phi)=0$) in each meson field;
\begin{align}
\Box& \sigma_i+ m_\sigma^2 \sigma_i + g_2\sigma_i^2+g_3\sigma_i^3 =
%{\color{red}\cancel{-}}
\hspace{1mm} g_\sigma \langle \bar{\psi}\psi\rangle_i, \\
\Box& \omega_{\mu,i} + m_\omega^2 \omega_{\mu,i} + c_3 (\omega_{\nu,i} \omega^{\nu}_i) \omega_{\mu,i} =\hspace{1mm} g_\omega \langle \bar{\psi}\gamma_\mu\psi\rangle_i, \\
\Box& \vec{\rho}_{\mu,i}+ m_\rho^2\vec{\rho}_{\mu,i} =\hspace{1mm} g_\rho \langle \bar{\psi}\vec{\tau}\gamma_\mu\psi\rangle_i, \\
\Box& A_{\mu,i} =\hspace{1mm} g_\omega \langle \bar{\psi}Q\gamma_\mu\psi\rangle_i,
\end{align}
where $\Box$ means d'Alembertian.
Note that they are integrated after multiplying the nucleon density of $i$-th particle, 
%{\color{red}
\begin{align}
\rho_i(\vec{r}) = \frac{1}{(2\pi L)^{3/2}}
e^{-\frac{(\vec{r}-\vec{r}_{i})^2}{2L}},
\end{align}
%}
that is, for the $\sigma$ meson,
\begin{align}
\sigma_i \equiv \int d\vec{r} \sigma(\vec{r}) \rho_i(\vec{r}). \end{align}
%\hspace{3mm}
Futher, we approximate as,
\begin{align}
\sigma_i^2 \approx \int d\vec{r} \sigma^2(\vec{r}) \rho_i(\vec{r}), \hspace{3mm}
\sigma_i^3 \approx \int d\vec{r} \sigma^3(\vec{r}) \rho_i(\vec{r}).
\end{align}
For the scalar density,
\begin{align}
\langle \bar{\psi}\psi\rangle_i \equiv \int d\vec{r} \sum_{j\neq i} f_j \rho_{j}(\vec{r}) \rho_{i}(\vec{r})
=  \sum_{j\neq i} f_{j} \rho_{ij},
\end{align}
where $\rho_{ij}$ is given by,
\begin{align}
\rho_{ij} =& \frac{\gamma_{ij}}{(4\pi L)^{3/2}} e^{-\frac{1}{4L}\tilde{\vec{R}}_{ij}^2}.
\label{grhoij}
\end{align}
We note that $\tilde{\vec{R}}_{ij}^2$ is the square of the distance between $i$-th and $j$-th nucleons,
\begin{align}
\tilde{\vec{R}_{ij}^2} =& (\vec{r}_i - \vec{r}_j)^2 + \gamma_{ij}^2
((\vec{r}_i - \vec{r}_j)\cdot\vec{\beta}_{ij})^2,
\label{rij}
\end{align}
%We note that this is different from Eq.(\ref{rhoij}) by factor
and 
$\gamma_{ij} = 1/\sqrt{1-\vec{\beta}_{ij}^2}$, where $\vec{\beta}_{ij} = (\vec{p}_i + \vec{p}_j)/(p_i^0 + p_j^0)$
is included due to the relativistic approach
\cite{Oliinychenko2016}.

%{\color{red} 
Similarly, 
we have,
\begin{align}
\omega_{\mu,i} \equiv & \int d\vec{r} \omega_\mu(\vec{r}) \rho_i(\vec{r}), \\ %\hspace{3mm}
\langle \bar{\psi}\gamma_\mu\psi\rangle_i \equiv & \sum_{j \neq i} \beta_{\mu,j}  \rho_{ij},  \hspace{3mm} \beta_{\mu,j} \equiv \frac{p_{\mu,j}}{\sqrt{\vec{p}_j^2+m_j^2}},
%(0, \vec{\beta}_j).
\end{align}
for the $\omega$ meson field,
%}
%{\color{red}
\begin{align}
\vec{\tau}_i\cdot\vec{\rho}_{\mu,i} &\equiv \int d\vec{r} 
\hspace{1mm}
\vec{\tau}\cdot\vec{\rho}_{\mu}
(\vec{r}) \rho_i(\vec{r}), \\
\vec{\tau}_i\cdot \langle \bar{\psi}\vec{\tau}\gamma_\mu\psi\rangle_i &\equiv \sum_{j\neq i} \tau_{3,i} \tau_{3,j} \beta_{\mu,j}\rho_{ij},
\end{align}
for the $\rho$ meson field, and,
\begin{align}
A_{\mu,i} \equiv & \int d\vec{r} A_\mu(\vec{r}) \rho_i(\vec{r}), \\ %\hspace{3mm}
\langle \bar{\psi}Q\gamma_\mu\psi\rangle_i \equiv & \sum_{j \neq i} 
Q_i Q_j\beta_{\mu,j}  \rho_{ij},
%(0, \vec{\beta}_j).
\end{align}
for the photon field.
Note that
the free mass and momentum are used
to estimate the scaler density and the baryon current
instead of those in nuclear medium, 
as done in Ref.~\cite{Nara2020}.
%}
%{\color{red}
%\sout{and we approximate as,}
%\begin{align}
%\cancel{
%(\omega_{\nu,i} \omega^{\nu}_i) \omega_{\mu,i} \approx \int %d\vec{r} (\omega_{\nu}(\vec{r}) \omega^{\nu}(\vec{r}))
%\omega_{\mu}(\vec{r}) \rho_i(\vec{r}).
%}\nonumber
%\end{align}
%}
In RQMD.RMF model, the derivatives of the meson field are neglected in the actual simulation\cite{Buss2012, Bernhard1991}, except for photon field, which is evaluated as the Coulomb interaction.
%(shown in the last term of Eq. (\ref{sky_pote})). 
%{\color{red}
%\sout{The expectation for the nucleon field is evaluated by assuming the no-meson field as \hbox{\cite{Nara2019, Nara2020}},}}
%{\color{red}
%\begin{align}
%\cancel{
%\langle \bar{\psi}\psi\rangle_i =\hspace{1mm} \sum_{j\neq i} f_j %\rho_{ij}, \nonumber
%}
%\\
%\cancel{
%\langle \bar{\psi}\gamma_\mu\psi\rangle_i =\hspace{1mm} \sum_{j\neq i} \beta_{\mu,j} \rho_{ij}, \nonumber
%}
%\\
%\cancel{
%\vec{\tau}_i\cdot \langle \bar{\psi}\vec{\tau}\gamma_\mu\psi\rangle_i =\hspace{1mm} \sum_{j\neq i} \tau_{3,i} \tau_{3,j} \frac{p_{\mu,j}}{p_j^0}\rho_{ij}. \nonumber
%}
%\end{align}
%}

To summarize the meson field equations in RQMD.RMF;
\begin{align}
\label{eq_sigma}
 m_\sigma^2 \sigma_i + g_2\sigma_i^2+g_3\sigma_i^3 =&\hspace{1mm} 
 %{\color{red}\cancel{-}} 
 g_\sigma \sum_{j\neq i} f_j \rho_{ij}, \\
\label{eq_omega}
m_\omega^2 \omega_{\mu,i} 
%{\color{red}
%\cancel{
%+ c_3 (\omega_{\nu,i} \omega^{\nu}_i) \omega_{\mu,i}}}
=&\hspace{1mm} g_\omega\sum_{j\neq i} \frac{p_{\mu,j}}{p_j^0} \rho_{ij}, \\
m_\rho^2 \vec{\tau}_i\cdot\vec{\rho}_{\mu,i} =&\hspace{1mm} g_\rho \sum_{j\neq i} \tau_{3,i} \tau_{3,j} \frac{p_{\mu,j}}{p_j^0}\rho_{ij},
\end{align}
and the photon field,
\begin{align}
eA_{\mu,i} =&\hspace{1mm} \alpha\hbar c 
\sum_{j\neq i} \frac{1}{|\tilde{\vec{R}}_{ij}|} {\rm erf}\left(\frac{|\tilde{\vec{R}}_{ij}|}{\sqrt{4L}}\right)
%{\color{red}
\delta_{\mu0}
%}
.
\end{align}
%{\color{red} \sout{
%On the other hand, the Euler-Lagrange equation gives the Dirac equation for nucleon field as,}
%\begin{align}
%\cancel{
%\left(
%\Slash{p} - \Slash{\Sigma}_v - (m_N - \Slash{\Sigma}_s) 
%\right)
%\psi = \left(
%\Slash{p}^* - m^* 
%\right)
%\psi = 0.\nonumber
%}
%\end{align}
%where $p_\mu^* = p_\mu - \Sigma_{\mu,v}$ and $m^* = m_N - \Sigma_{s}$.
%\sout{
%The scalar potential, $\Sigma_{s}$ and vector potential, $\Sigma_v^{\mu}$ can be represented as,}
%\begin{align}
%\cancel{
%\Sigma_{s} =\hspace{1mm} g_\sigma \sigma, } \nonumber\\
%\cancel{
%\Sigma_{v}^\mu =\hspace{1mm} g_\omega \omega^\mu + g_\rho %\vec{\tau}\cdot\vec{\rho}^\mu  + e A^\mu,} \nonumber
%\end{align}
%\sout{
%respectively. In the Hamiltonian, we have}
%\begin{align}
%&\sum_i
%int d\vec{r}
%\bar{\psi}_i
%\left(
%-i \Slash{\vec{\nabla}} + m_N + g_\sigma \sigma + g_\omega %\Slash{\omega} + g_\rho \vec{\tau}\cdot \Slash{\vec{\rho}} + eQ \Slash{A} 
%\right)
%\psi_i = \sum_i e_i, \\
%\cancel{
%H =  \sum_i e_i,} \nonumber \\
%\cancel{
%e_i = p^0_i +  \Sigma_{v}^0 = \sqrt{{\vec{p}_i}^2 + {m_N}^2} + \Sigma_{v}^0.} \nonumber
%\end{align}
%}
In the actual simulation, $\sigma_i$ 
%{\color{red}\sout{and $\omega^\mu_i$}} 
needs to be solved iteratively via Eq.~(\ref{eq_sigma})%{\color{red}\sout{ and (\ref{eq_omega})}}
.
If the momentum dependent interaction is considered in RQMD.RMF model, they may be added in the above potentials as\cite{Nara2020},
\begin{align}
\Sigma_{s,i} &=\hspace{1mm} g_\sigma \sigma_i + V_{s,i}^{\rm MD}, 
\\
\Sigma_{v,i}^\mu &=\hspace{1mm} g_\omega \omega_i^\mu + g_\rho \vec{\tau}_i\cdot\vec{\rho}_i^\mu  + e A_i^\mu + V_{v,i}^{\mu,{\rm MD}},
\end{align}
where
\begin{align}
V_{s,i}^{\rm MD} &=\hspace{1mm} \frac{\bar{g}_\sigma^2}{m_\sigma^2} \sum_{j\neq i} f_j F_{ij}^{\sigma} \rho_{ij}, \\
V_{v,i}^{\mu,{\rm MD}} &= \hspace{1mm}
\frac{\bar{g}_\omega^2}{m_\omega^2}\sum_{j\neq i} \frac{p_j^\mu}{p_j^0} F_{ij}^{\omega} \rho_{ij}
+\frac{\bar{g}_\rho^2}{m_\rho^2}\sum_{j\neq i} \frac{p_j^\mu}{p_j^0} F_{ij}^{\rho} \tau_{3,i} \tau_{3,j} \rho_{ij},
\end{align}
where the momentum dependent term is neglected in the photon field. $F_{ij}^{\rm meson}$ has a Lorentzian-type momentum dependent form;
\begin{align}
F_{ij}^{\rm meson} = \frac{1}{1+\tilde{\vec{p}}_{ij}^2/\Lambda_{\rm meson}^2},
\end{align}
where $\Lambda_{\rm meson}$ is the cutoff energy.
%{\color{red}
This study did not include the momentum dependent terms, because the relatively low incident energy is considered for hadron therapy application.
%}
The parameters in 
%{\color{red}\sout{RMF} 
the non-linear sigma model
%} 
are $g_\sigma$, $g_2$, $g_3$, $g_\omega$, $g_\rho$,
%{\color{red}\sout{
%$\bar{g}_\sigma$, $\bar{g}_\omega$, $\bar{g}_\rho$,
%$\Lambda_\sigma$, $\Lambda_\omega$, $\Lambda_\rho$,}}
$m_\sigma$, $m_\omega$, and $m_\rho$.
The parameter sets employed in this study can be found in
Appendix.
%Ref.\cite{PhysRevC.102.024913}
%, though $\rho$-meson (and photon?) was not considered there.

%The end of this subsection, the evaluation of the self-energy of meson field is noted.
Finally, we note how the self-energy of the meson field is evaluated.
We have the following relation in sigma meson energy,
\begin{align}
\int d\vec{r} \left(
m_\sigma^2 \sigma^2
%{\color{red}
(\vec{r})
%}
+ g_2\sigma^3
%{\color{red}
(\vec{r})
%}
+g_3\sigma^4
%{\color{red}
(\vec{r})
%}
\right) = \sum_i \Sigma_{s,i} f_i,
\label{selfenergy_sigma}
\end{align}
where $m_\sigma^2 \sigma^2
%{\color{red}
(\vec{r})
%}
+ g_2\sigma^3
%{\color{red}
(\vec{r})
%}
+g_3\sigma^4
%{\color{red}
(\vec{r})
%}
= \sigma
%{\color{red}
(\vec{r})
%}
g_\sigma
\sum_{i} f_i \rho_{i}
%{\color{red}
(\vec{r})
%}
$ is applied. The left-hand side in (\ref{selfenergy_sigma}) is now approximated as,
\begin{align}
%\int d\vec{r} \left(
%m_\sigma^2 \sigma^2 + g_2\sigma^3+g_3\sigma^4 \right) 
{\rm LHS\hspace{1mm}in\hspace{1mm}(\ref{selfenergy_sigma})}
\approx 
\sum_i \left(
m_\sigma^2 \sigma_i^2 + g_2\sigma_i^3+g_3\sigma_i^4 \right)\Delta r_i.
\label{dr_0}
\end{align}
From Eqs.~(\ref{selfenergy_sigma}) and (\ref{dr_0}), we obtain,
\begin{align}
\Delta r_i = \frac{\Sigma_{s,i} f_i}{m_\sigma^2 \sigma_i^2 + g_2\sigma_i^3+g_3\sigma_i^4}.
\label{selfenergy_sigma_r}
\end{align}
Thus, we evaluate the self energy of $\sigma$ meson as,
\begin{align}
&\int d\vec{r} 
\left(
- \frac{1}{6}g_2 \sigma^3 - \frac{1}{4}g_3 \sigma^4
\right) \nonumber \\
%\approx & 
% \sum_i
% \left(
% (- \frac{1}{6}g_2 \sigma_i^3 - \frac{1}{4}g_3 \sigma_i^4) \Delta r_i
% \right)\\
&\approx
 \sum_i
 \left(
(- \frac{1}{6}g_2 \sigma_i - \frac{1}{4}g_3 \sigma_i^2)
\frac{\Sigma_{s,i} f_i}{m_\sigma^2 + g_2\sigma_i + g_3\sigma_i^2}
\right).
\end{align}
%{\color{red}\sout{
%Similalily, the self energy of the omega meson is given as,}
%\begin{align}
%\cancel{
%\int d\vec{r} 
%\left(
%\frac{1}{4}c_3 (\omega_\mu\omega^\mu)^2 } \nonumber\\
%\right) 
%= & 
% \sum_i
% \left(
% (- \frac{1}{6}g_2 \sigma_i^3 - \frac{1}{4}g_3 \sigma_i^4) \Delta r_i
% \right)\\
%\cancel{
%\approx
% \sum_i
% \left(
%\frac{1}{4}c_3 (\omega_{\mu,i}\omega^{\mu}_i)
%\frac{g_\omega\beta_{\mu,i}\omega^\mu_i}
%{m_\omega^2 + c_3(\omega_{\nu,i}\omega^\nu_i)^2}
%\right).}\nonumber
%\end{align}
%}

\subsection{\label{sec:GSP}Evaluation of ground state properties}
The RQMD.RMF model is first evaluated based on the ground state properties of 10 stable nuclides ($^{12}$C, $^{14}$N, $^{16}$O, $^{31}$P, $^{40}$Ca, $^{63}$Cu, $^{90}$Zn, $^{120}$Sn, and $^{208}$Pb). In this study, (1) the charge radius, (2) stability, and (3) the relation between binding energy and nuclear size 
of the ground states calculated with RQMD.RMF are evaluated.
%are evaluated as ground state properties within the RQMD.RMF model. 
The most suitable parameter set is selected based on this evaluation and subsequently applied to the fragmentation analysis.

\subsubsection{Root mean squared charge radius}
The root mean squared (RMS) charge radius of ground state nuclei is defined as
\begin{align}
R_c &= \sqrt{\langle r^2 \rangle}, \label{R_c}\\
\langle r^2 \rangle &= \int r^2 \rho_c (\vec{r}) d\vec{r},
\end{align}
where $\rho_c (\vec{r})$ is calculated by considering only protons immediately after the binding energy adjustment in the QMD simulation. This result is then compared with the experimental data\cite{Angeli2013}.
\subsubsection{Stability}
The stability of the ground-state nucleus produced by the QMD model is evaluated by monitoring the fluctuation of binding energy within 100 [fm/c]. Specifically, the absolute difference between the maximum and minimum binding energy during time evolution is recorded for each parameter set.
\subsubsection{Binding energy and nuclear size}
The binding energy of a nucleon cluster sampled from, 
\begin{align}
\rho_{ws}(\vec{r}) = \frac{\rho_0}{1+\exp{(-(r-r_0)/a_{ws})}} Y_{00}(\theta,\phi),
\label{WS}
\end{align}
where the radial part has the form of Wood-Saxon profile with $r_0 = 1.124 A^{1/3} - 0.5$ [fm] and $a_{ws}=0.2$ [fm], is evaluated. $\rho_0$ is the normalization factor.
Here, if the absolute difference from the experimental binding energy is less than 0.5 MeV/A, the sampling is considered successful for the construction of the ground-state nucleus.
The number of sampling trials required for success can be regarded as an indicator of the efficiency in creating a ground-state nucleus with the proper binding energy and nuclear size.
%The sampling of nucleons based on the Wood-Saxon model statistically can generate the proper nuclear profile. If the binding energy with such the sampling is reproducible with the experimental data, the nuclear potential is considered to be proper. 

\subsection{\label{sec:FragmentEx}Fragmentation cross section analysis}
The performance of the parameter set selected based on the above ground-state properties is then assessed using fragment cross-sections in a $^{12}$C ion beam experiment conducted at GANIL (Accélérateur National d’Ions Lourds) and QST (National Institutes for Quantum Science and Technology).
\subsubsection{GANIL data}
In the GANIL experiment, fragment production cross-sections, differential cross-sections (angular distributions), and double-differential cross-sections (energy distributions for specific angles) were measured using 50 and 95 MeV/u carbon ion beams incident on targets of C, CH$_2$, Ti, Al, Al$_2$O$_3$, and PMMA (C$_5$H$_8$O$_2$) \cite{Dudouet2013, Divay2017}.
Sixteen fragments were identified for targets of $^1$H, $^{12}$C, $^{16}$O, $^{27}$Al, and $^{\rm nat}$Ti:
$^1$H, $^2$H, $^3$H, $^3$He, $^4$He, $^6$He, $^6$Li, $^7$Li, $^7$Be, $^9$Be, $^{10}$Be, $^8$B, $^{10}$B, $^{11}$B, $^{10}$C, and $^{11}$C.
For 95 MeV/u, the $^{12}$C fragment was also included. Each fragment was measured with a specific energy threshold \cite{Dudouet2014, Divay2017}.

In the present study, $^1$H, $^{12}$C, $^{16}$O, $^{27}$Al, and $^{48}$Ti are considered as targets, and the absolute difference of angular distribution from the experimental data
is evaluated in each target.

\subsubsection{QST data} 
Charge-changing cross-sections and fragment production cross-sections were measured using 290 and 400 MeV/u carbon ion beams incident on targets of C, CH$_2$, Al, Cu, Sn, and Pb \cite{Zeitlin2007}.
For fragment production, cross-section results were provided for three acceptance angles (2.5, 3.9, and 7.3 degrees) for effective charges that could be distinguished in the experiment.

In this study, $^1$H, $^{12}$C, $^{27}$Al, and $^{63}$Cu are considered as targets. The production cross-sections for the effective charge values
$Z_{eff} \in (5, 4.4, 4, 3.5, 3, 2, 1)$
at the three acceptance angles mentioned above
are derived from the QMD simulation.
The absolute difference between the charge-changing cross-section and the fragment production cross-sections for all three acceptance angles and the experimental data is evaluated for each target.

\subsection{\label{sec:QMDpara}Parameters in QMD}
Our code is developed based on the Light-Ion QMD (LIQMD) model, which employs the Skyrme force and includes an ad hoc $\alpha$-cluster model for light nuclei, implemented in Geant4 ver.~11.2 \cite{Sato2022}.
The LIQMD model has been optimized for two parameters: the square of the width of the Gaussian wave packet, denoted as $L$ in Eq.~(\ref{grhoij}), and the maximum distance criterion in the fragment cluster judgment, denoted as $R$,
\begin{align}
\tilde{\vec{R}_{ij}^2} 
%= (\vec{r}_i - \vec{r}_j)^2 + \gamma_{ij}^2((\vec{r}_i - \vec{r}_j)\cdot\vec{\beta}_{ij})^2
\le R^2,
\label{clusterR}
\end{align}
for carbon ions incident at 400 MeV/u on a water target using a multi-objective optimization technique.
The LIQMD model 
employs the SkM$^*$ Skyrme nucleon-nucleon interaction\cite{Bartel1982}.
Therefore, $L$ and $R$ used in Geant4 should be recognized as unique values within that interaction.
While $L$ and $R$ could, in principle, be optimized independently for each RMF parameter set employed in this study, doing so would complicate the simulation process, including fragmentation analysis.
Therefore, in this study, $L$ is optimized by using a single parameter set--one of the latest parameter sets proposed for finite nucleus analysis, NL3$^*$\cite{Lalazissis1997}--for the RMS radii, and same $L$ is applied to other parameter sets.
In this optimization, the $\alpha$ cluster parameters ($(a,b)$ of Eqs.~(6), (8) and (9) in Ref.\cite{Sato2022}) are determined simultaneously. As a result, we obtain
\begin{align}
L = 0.035A^{2/3}+0.9, \hspace{3mm} (a,b) = (0.7, 3.0),
\end{align}
in the present study using the RQMD.RMF model.
In nucleus-nucleus collision, the value of $L$ is set as the weighted mean of the two nuclei.

In contrast to $L$, the clustering distance $R$ can be optimized through fragmentation analysis, as $R$ directly affects to the fragment formation. However, in this study, $R = 4$ [fm], a typical value in the clustering radius for QMD simulation\cite{Zhang2012}, is employed in the following simulation. Although the clustering radius may fundamentally depend on the nucleon-nucleon interaction, it is reasonable to assume a similar radius across different nuclear models.

%In this process, $R$ has an isospin-dependency\cite{Zhang2012}, that is, $R$ with same isospin (here after $R_{pp}$ and $R_{nn}$ denote the proton-proton and neutron-neutron clustering distance, respectively) and $R$ with different isospin ($R_{pn}$, the proton-neutron clustering distance) are separately optimized. As a tuning result, we find that $R_{pp}$ and $R_{pn}$ are not very sensitive to the incident kinetic energy and the species of the participant nucleus in fragment productions, whereas $R_{pn}$ can weakly depend on them. In this study, we set $R_{pp} = R_{nn} = 4$ [fm] and, 
%\begin{align}
%R_{pn} =& 10.6 - 0.00331\bar{A} - 0.00016\bar{A}^2 \nonumber \\
%& - 1.09 e_k - 0.00048 e_k^2 - 5.85\times10^{-7} e_k^3 \\
%& + 0.000114 e_k \bar{A} \hspace{2mm}{\rm [fm]}, \nonumber
%\end{align}
%where $\bar{A}$ is the mean of the atomic mass number in two participant nuclei, and $e_k$ is the incident kinetic energy [MeV].
%If $R_{pn} > $  7 ($R_{pn} < $  4), it is replaced with $R_{pn}$ = 7 ($R_{pn}$ = 4).

%{\color{red}\sout{Two} 
Three
%} 
additional parameters inherent to the QMD model should be mentioned here:
%{\color{red}
the time step in time evolution ($dt$)
%}
,
the maximum simulation time for inelastic collisions within QMD ($T_{max}$), and the maximum sampling radius for the impact parameter ($b_{max}$).
%{\color{red} 
Our code has been developed using the Geant4 toolkit, which employs the second-order Runge–Kutta (RK) method with a time step of 
$dt = 1$ [fm/c] in the QMD dynamics simulation.
In this study, we adopt the same settings, except for the monitoring of the binding energy fluctuation in the ground state nucleus, for which the fourth-order RK method is also applied to confirm that the fluctuation originates from numerical integration.
%}

In our code, the participant nuclei (projectile and target) are initially set at a distance such that they will collide after $t_{init} = 20$ [fm/c]. However, if the distance is less than a radius calculated from the reaction cross-section, $b_{\sigma} = \sqrt{\sigma_R/\pi}$, where $\sigma_R$ is the reaction (inelastic) cross-section, the initial distance is replaced with $b_{\sigma}$. This implies that the initial distance between the projectile and target depends on the incident kinetic energy.
The maximum simulation time $T_{max}$ is defined as $T_{max} = t_{init} \times a_{tmax}$ [fm/c], where $t_{init}$ represents the time required to traverse the initial distance and $a_{tmax} = 3$ is employed in this study ($T_{max} = 60$ unless the distance is less than $b_{\sigma}$). 
%depends on $\bar{A}$ and $e_k$. 
%From GANIL and QST data, we set $a_{tmax}$ as, 
%\begin{align}
%a_{tmax} =& \frac{4}{1+\exp{(-0.00874\bar{A} - 0.00425e_k)}},
%\end{align}
%meaning 2 $< a_{tmax} <$ 4.

In simulating nucleus–nucleus collisions, the impact parameter (commonly referred to as the 
$b$-value) is initially sampled within a specific range. Here, the impact parameter is determined by multiplying
$b_{max} = b_{env}b_{\sigma}$ with the square root of a uniform distribution in the range [0,1]), where 
$b_{env}$ is an extension beyond the radius determined by $b_{\sigma}$
to account for the statistical reaction probability in peripheral collisions. 
In this study, $b_{env} = 1.2$ for a mean mass 
of projectile and target,
$\bar{A} \le 20$, while $b_{env} = 1.05$ is applied otherwise. 

%except for H target, where $b_{env}$ is changed as 1.6, 1.7, 1.1, and 1.0 for kinetic energy of 50, 95, 290, and 400 MeV/u, respectively. 

%For above parameter tuning, we simulate with $10^5$ events for corresponding GANIL and QST experiments, where the parameter grids of $R_{pn} \in (4, 5, 6, 7)$ [fm],  $a_{tmax} \in (2.5, 3.0, 3.5)$, and $b_{env} \in (1.1, 1.2, 1.3, 1.4, 1.5, 1.6, 1.7)$
%are employed.
%This study models 
%$b_{env}$ as a function of $\bar{A}$ and $e_k$.

\subsection{Collision}
The dynamics of participant nucleons is simulated using 
Eqs.~(\ref{eomRQMD1}) and (\ref{eomRQMD2}). However, the interactions in QMD models are insufficient to generate a hardcore potential, which is considered crucial for realistic nucleon-nucleon collision phenomena. Moreover, these interactions do not account for nucleon excitations or their sequential decay into mesons, such as pions. To address this, nucleon-nucleon (as well as meson-nucleon and meson-meson) collisions, allowing both elastic and inelastic processes, are independently incorporated into the time evolution of the QMD model. The included channels in the present model are same as those in Binary Cascade model implemented in Geant4 ($\mathtt{G4Scatterer}$). Details are found in the references\cite{Folger2004}.
Because $\mathtt{G4Scatterer}$ is modeled for free particles, the scattering in our model includes the Pauli's exclusion phenomenon for the nuclear medium, as the scattered nucleons occupy a state allowed by Fermi statistics. 

\subsection{Fragmentation}
Finally, we mention the treatment of the fragment formation after the time evolution. 
After the termination of the time evolution in the reaction system, a cluster identification is carried out.
Each identified cluster is considered as a fragmented nucleus from the reaction and it usually has an excitation energy, which is evaluated within the QMD model.
Sequentially, the excited nucleus is de-excited.
In our code, the de-excitation is handled by $\mathtt{G4ExcitationHandler}$, where three processes are mainly considered\cite{Quesada2011}; 
\begin{itemize}
\item Multi-fragmentation
\item Fermi breakup
\item Evaporation
\end{itemize}
If the identified cluster is unstable and the excitation energy $E-E_{ground} > 3A$ MeV, where $E$ is the energy of identified cluster calculated by QMD while $E_{ground}$ is the experimental energy in the ground state,
a statistical multi-fragmentation\cite{Bondorf1995} take place once.
The produced fragments for $Z \le 8$ and $A \le 16$ are then de-excited by Fermi breakup model\cite{Bondorf1995}.
The processes which are not considered by Fermi breakup model are included in evaporation model.
$\mathtt{G4ExcitationHandler}$ uses the generalized evaporation model (GEM)\cite{Furihata2001} which considers up to $^{28}$Mg.

\section{\label{sec:result}Result}
\subsection{Ground-state properties}
%{\color{red} 
The 
%\sout{MAEs} 
differences in the RMS charge radii for 10 stable nuclei between QMD simulations and experimental data are shown in Fig.~1. The mean value and standard deviation for each parameter set are calculated from 100 trials.
%}
%The \sout{MAEs} difference} of the RMS charge radius for 10 stable nuclei obtained from {\color{red} between QMD simulations and experimental data} are shown in Fig.~\ref{fig:charge}, where the mean value and standard deviation for each parameter set are calculated from 100 trials.
The charge radii are reproduced within 0.2 fm across a wide mass range. Although nuclear size can be sensitive to the width of the Gaussian wave packet, which is optimized in the NL3$^*$ parameter set, this trend in charge radius reproducibility suggests that there is no significant dependency of $L$ among the RMF parameter sets in generating a ground-state nucleus with a reasonable size.

\begin{figure}[h]
\includegraphics[width=9cm]{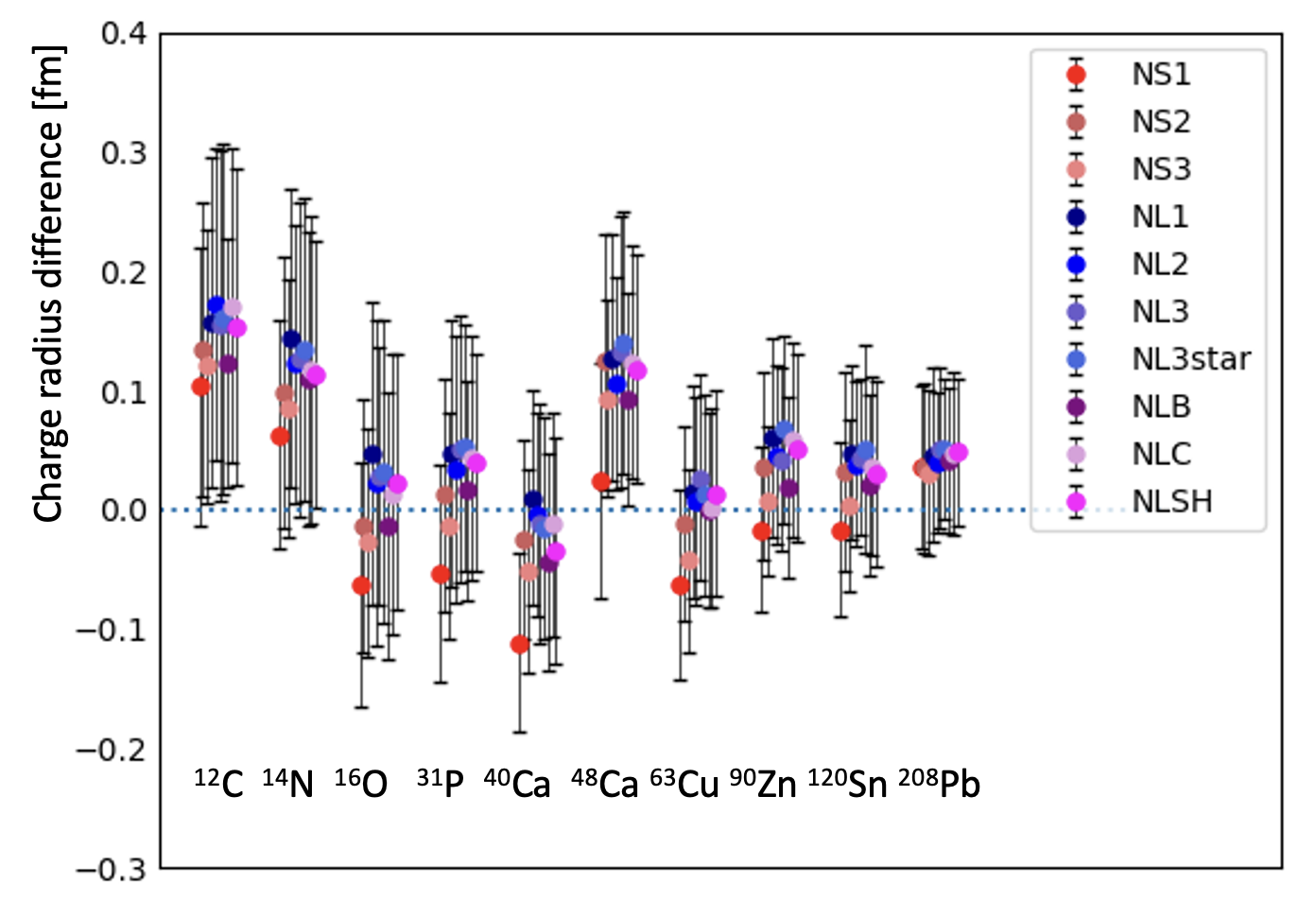}
% Here is how to import EPS art
\caption{\label{fig:charge} Difference in charge radius between the results calculated using Eq.~(\ref{R_c}) and experimental data \cite{Angeli2013}. The mean value and standard deviation from QMD simulations for 10 RMF parameter sets are shown.}
\end{figure}

\begin{figure}[h]
\includegraphics[width=9cm]{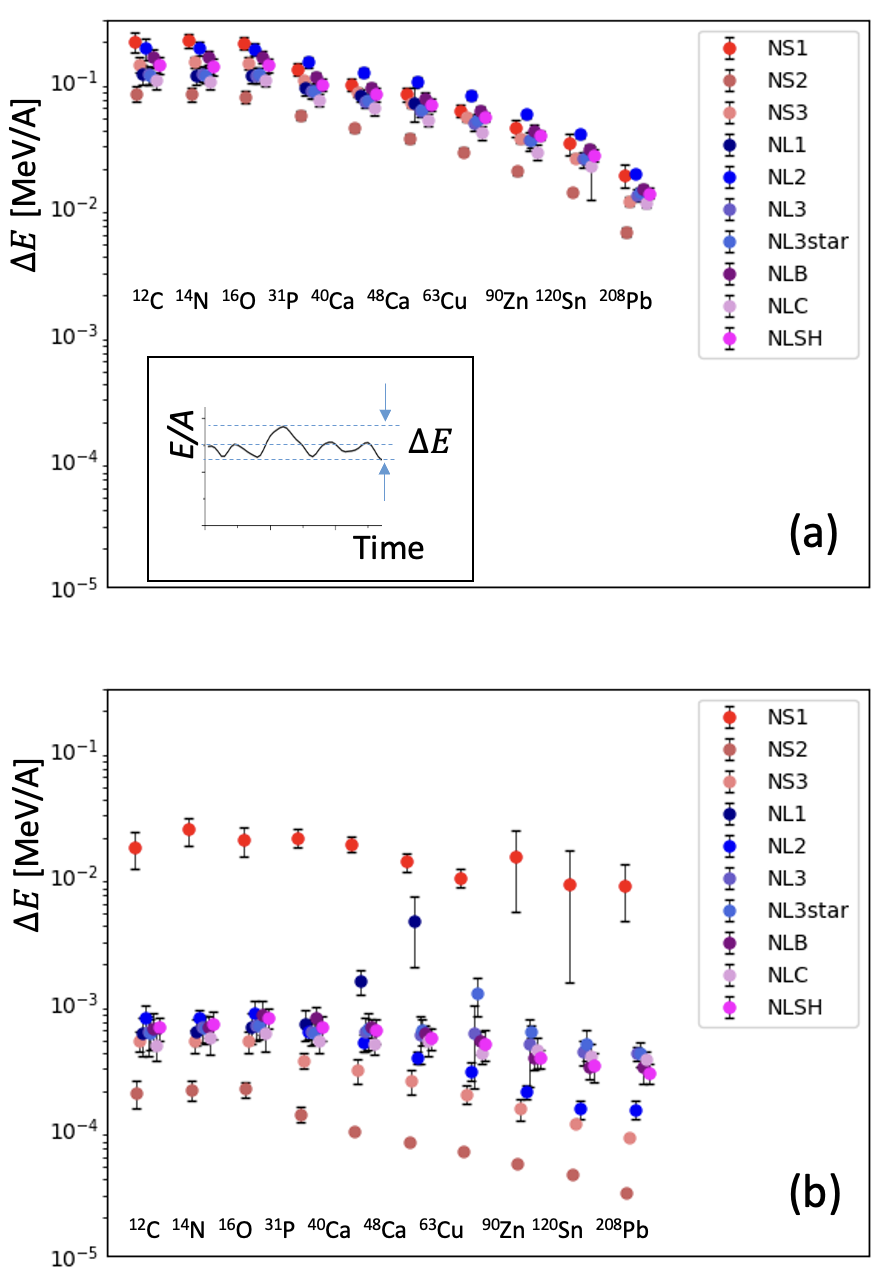}% Here is how to import EPS art
\caption{\label{fig:stability} Stability analysis by monitoring the binding energy per nucleon, where the binding energy fluctuation is monitored as illustrated in the inset 
%{\color{red}
of the upper panel.
(a) Second-order Runge–Kutta method, and (b) fourth-order Runge–Kutta method, both with a time step of 1 [fm/c] in the dynamics simulation.
%}
The mean value and standard deviation of the absolute difference between the maximum and minimum energies are indicated for 
%{\color{red}\sout{16} 
10
%} 
RMF parameter sets. 
%{\color{red}\sout{Data points outside the displayed range are not shown.}
}
%}
\end{figure}

Figure \ref{fig:stability} illustrates the stability of the created nucleus, defined as the absolute difference between the maximum and minimum binding energies over 100 fm/c.
The mean values (circles) and standard deviations (bars) from 100 trials are plotted. 
%{\color{red} 
Figure \ref{fig:stability}(a) shows the result obtained using the second-order Runge–Kutta method with a time step of 1 [fm/c] in the dynamics simulation, whereas Fig.~\ref{fig:stability}(b) presents the result obtained using the fourth-order Runge–Kutta method. It is observed that the RK4 method reduces the energy fluctuation $\Delta E$, indicating that 
the main source of the binding energy fluctuation is the numerical integration.
%the source of the binding energy fluctuation arises from numerical integration errors.
%}
%{\color{red}\sout{Several parameter sets (MD1–MD4) exceed the vertical axis range of this figure, indicating that the ground-state nucleus created by the RQMD.RMF model is relatively unstable with these parameters.}} 
The dynamics simulation in the RQMD.RMF model is more stable for heavier nuclei. The NS2 parameter set yields the most stable results across the entire mass range.

\begin{figure}[h]
\includegraphics[width=9cm]{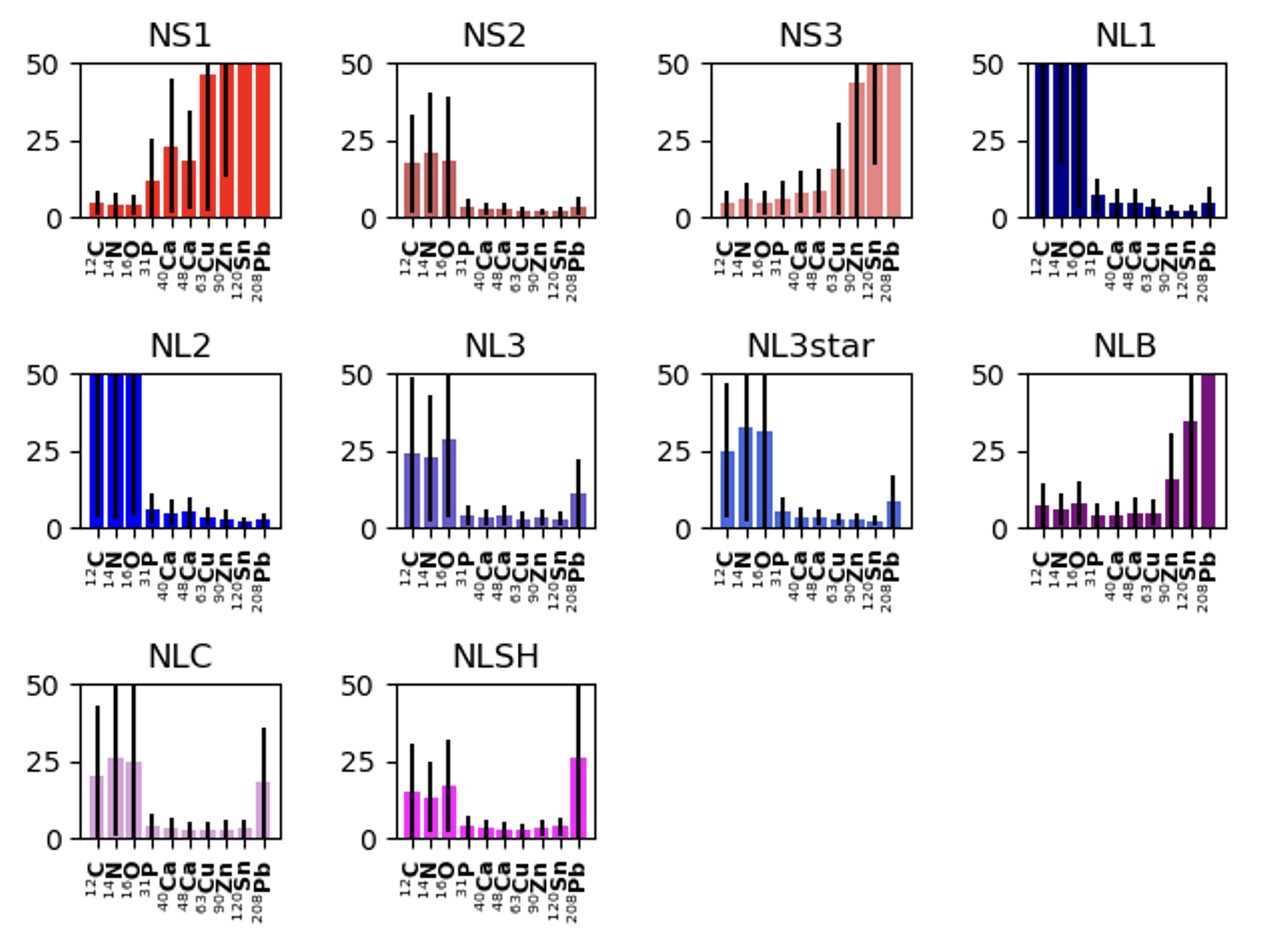}% Here is how to import EPS art
\caption{\label{fig:repeat} Number of sampling trials required to obtain a binding energy within 0.5 MeV/A of the experimental value.}
\end{figure}

The number of sampling trials required to obtain a binding energy within 0.5 MeV/A of the experimental data is shown in Fig.~\ref{fig:repeat}.
Since sampling is performed using the Wood-Saxon profile Eq.~(\ref{WS}), a lower number of trials implies that the binding energy is more easily reproduced with this empirical nuclear profile. From this perspective, it can be said that the NS2, NL3, NL3$^*$, NLC, NLSH
%{\color{red}\sout{, TM1, and PK1}} 
parameter sets allow the empirical Wood-Saxon profile and the experimental binding energy to coexist across a wide mass range. 
%{\color{red}
Relatively large trials for light nuclei may indicate that a more realistic profile,
such as a Gaussian-like profile, is required
for describing light nuclei.
%}

In terms of nuclear matter properties (Table II in the Appendix), identifying common features among these parameter sets may be challenging. However, relatively low incompressibility might be favorable for reproducing the binding energy with a reasonable nuclear profile in QMD simulations.

Based on the ground-state properties evaluated by QMD simulations, NS2 is selected as the parameter set for the subsequent fragmentation analysis. However, it should be noted that this selection does not imply that NS2 is the best parameter set. In this study, 
$L$ and the parameters in the $\alpha$-cluster model were determined using NL3$^*$. Additionally, while NS2 appears to be the most stable parameter set, this stability may not be a critical factor in the sequential analysis of fragment production.
The present results do not preclude further investigations using other parameter sets in RQMD.RMF simulations.

%\begin{figure*}
%\includegraphics{Charge_radius.png}% Here is how to import EPS art
%\caption{\label{fig:wide}Use the figure* environment to get a wide
%figure that spans the page in \texttt{twocolumn} formatting.}
%\end{figure*}

\subsection{Fragment production cross-sections}
\begin{figure}[h]
\includegraphics[width=9cm]{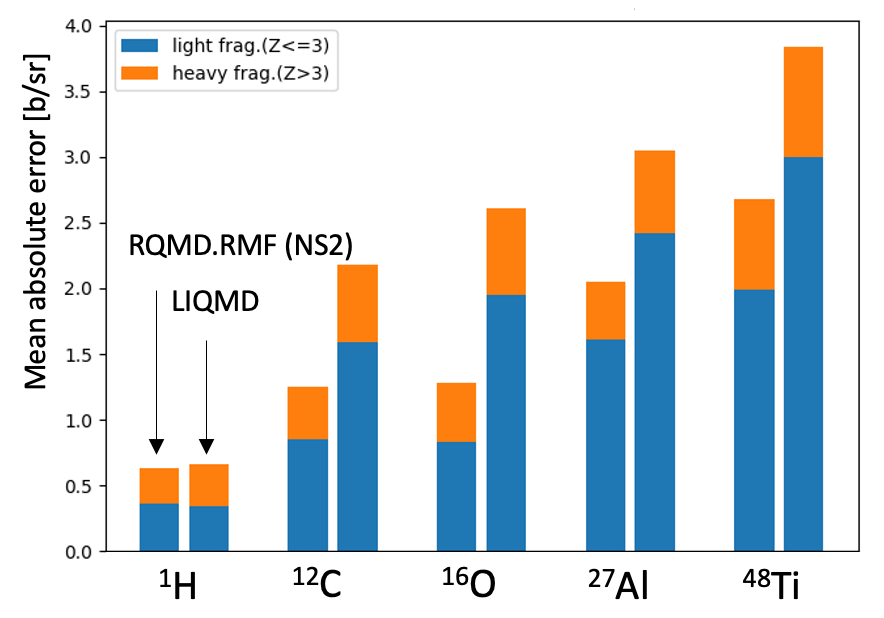}% Here is how to import EPS art
\caption{\label{fig:MAE50MeV} Mean absolute error of the differential cross-section in H, C, O, Al, and Ti targets for a 50 MeV/u carbon ion projectile. The left bar plots represent the results from the RQMD.RMF model using the NS2 parameter set. For comparison, the results from the LIQMD model implemented in Geant4 ver.~11.2 are shown as the right bars.}
\includegraphics[width=9cm]{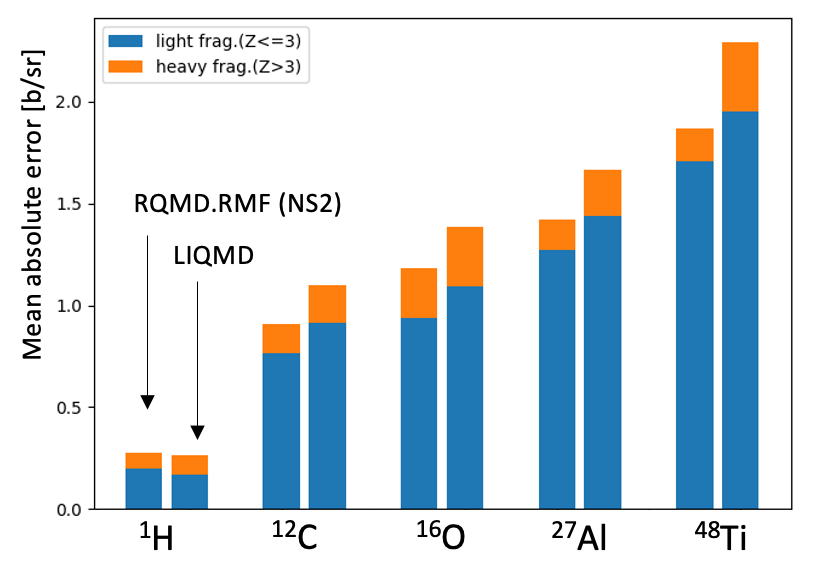}% Here is how to import EPS art
\caption{\label{fig:MAE95MeV}  Mean absolute error of the differential cross-section in H, C, O, Al, and Ti targets for a 95 MeV/u carbon ion projectile. The format is the same as in Fig.~\ref{fig:MAE50MeV}.}
\end{figure}

Fragment production cross-sections 
for carbon ion projectile at 50, 95, 290, and 400 MeV/u
are calculated using the RQMD.RMF model with NS2 parameter set.
The MAEs of angular distribution
in H, C, O, Al, and Ti targets 
for 50 MeV/u and 95 MeV/u carbon beams\cite{Dudouet2013, Divay2017}
are depicted in Figs.\ref{fig:MAE50MeV} and \ref{fig:MAE95MeV}, respectively.
In these figures, blue represents the MAE for lighter fragments ($^1$H, $^2$H, $^3$H, $^3$He, $^4$He, $^6$He, $^6$Li, and $^7$Li), whereas orange represents the MAE for heavier fragments ($^7$Be, $^9$Be, $^{10}$Be, $^8$B, $^{10}$B, $^{11}$B, $^{10}$C, and $^{11}$C. For 95 MeV/u, $^{12}$C is also included).

For comparison, results from simulations using the LIQMD model implemented in Geant4 ver.~11.2 (SkM$^*$ Skyrme interaction model) are also included. These figures show that the RQMD.RMF model generally outperforms the LIQMD model for all targets, except for the 95 MeV/u beams on H targets.
The MAE from lighter fragments is dominant, as the production yields of these fragments are significantly higher than those of heavier fragments. The angular distributions presented in the Appendix indicate that the RQMD.RMF model accurately describes angular distributions, even for heavier fragments (see Figs.~\ref{fig:50Angle} and \ref{fig:95Angle}).

\begin{figure}[h]
\includegraphics[width=9cm]{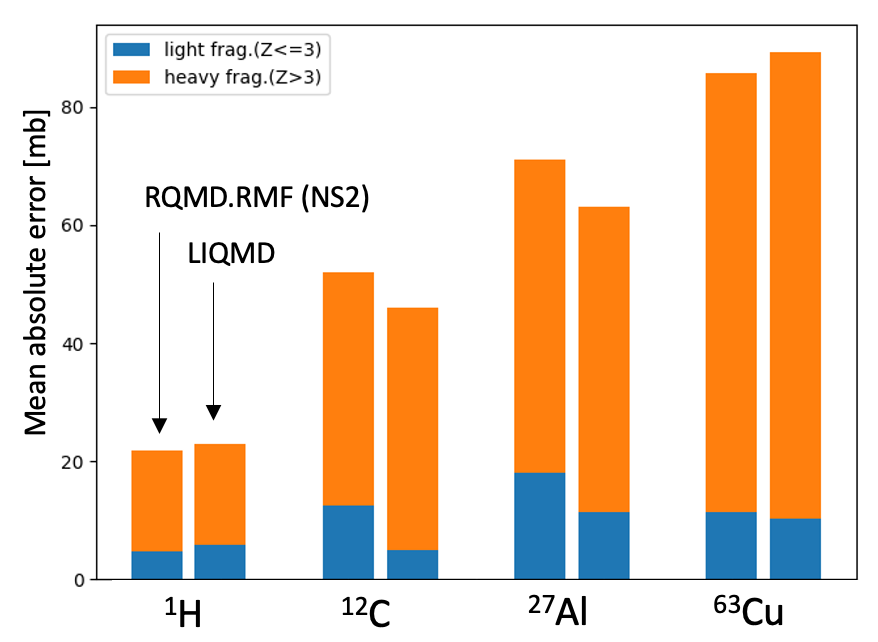}% Here is how to import EPS art
\caption{\label{fig:MAE290MeV} Mean absolute error of the production cross-section in H, C, Al, and Cu targets for a 290 MeV/u carbon ion projectile. The left bar plots represent the results from the RQMD.RMF model using the NS2 parameter set. For comparison, the results from the LIQMD model implemented in Geant4 ver.~11.2 are shown as the right bars.}
\includegraphics[width=9cm]{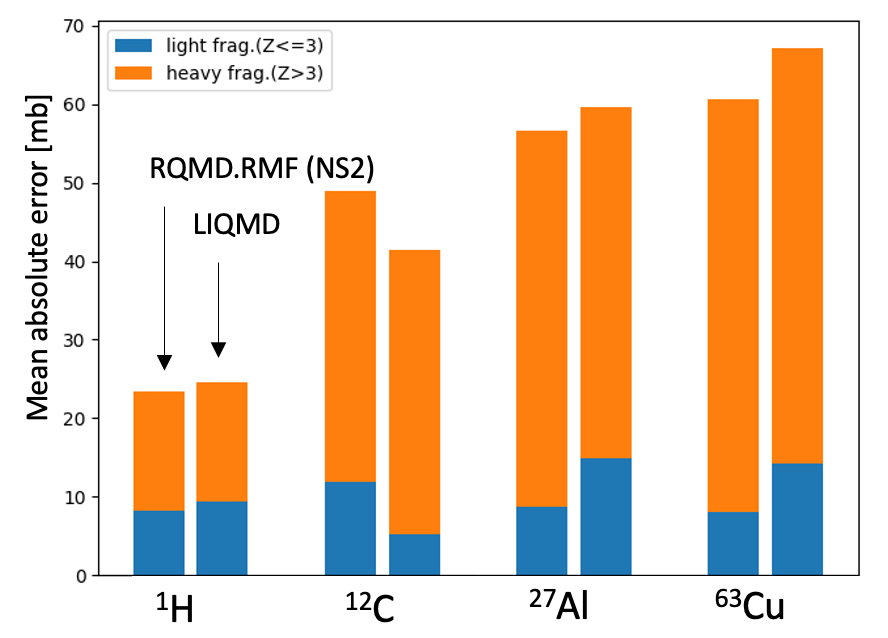}% Here is how to import EPS art
\caption{\label{fig:MAE400MeV}  Mean absolute error of the production cross-section in H, C, Al, and Cu targets for a 400 MeV/u carbon ion projectile. The format is the same as in Fig.~\ref{fig:MAE290MeV}.}
\end{figure}

The MAEs of the charge-changing cross-section and the production cross-section for three limited acceptance angles in H, C, Al, and Cu targets for 290 MeV/u and 400 MeV/u carbon beams are depicted in Figs.~\ref{fig:MAE290MeV} and \ref{fig:MAE400MeV}, respectively.  In these figures, blue represents the MAE for lighter fragments ($Z_{eff} \in (3, 2, 1)$), whereas orange represents the MAE for heavier fragments  ($Z_{eff} \in (5, 4.4, 4, 3.5)$).
These figures indicate that the RQMD.RMF model is comparable to the LIQMD model in reproducing the experimental data. Unlike the results for 50 and 95 MeV/u, the MAE for lighter fragments is smaller than that for heavier fragments at 290 and 400 MeV/u.
This is due to the relatively small acceptance angles (at most 7.3 degrees) considered in the 290 and 400 MeV/u beam experiments at QST. Specifically, the experimental data are weighted toward the forward direction. Light fragments have a non-negligible yield even at large polar angles; therefore, it can be inferred that the MAE at 290 and 400 MeV/u, as analyzed in this study, is more influenced by the production cross-section of heavier fragments.

\section{\label{sec:discussion}Discussion}
This study demonstrates that 
%{\color{red}\sout{16} 
10
%} 
parameter sets previously proposed within the framework of RMF theory are capable of reproducing experimental values using the QMD model for the ground-state properties of stable nuclei, specifically the RMS charge radius and binding energy per nucleon.
For the charge radius, lighter nuclei ($^{12}$C and $^{14}$N) and the neutron-rich nucleus ($^{48}$Ca) tend to overestimate the charge radius (Fig.\ref{fig:charge}). In particular, lighter nuclei were also found to be relatively unstable in the stability analysis, which monitors the binding energy over a 100 fm/c time evolution (Fig.~\ref{fig:stability})

Most parameter sets in the RMF model are typically determined using nuclear matter properties and evaluated in finite systems within a mean-field approach. Therefore, it is natural that RMF parameter sets describe heavier nuclei more accurately than lighter ones. However, previous studies have reported that the RMF model can reproduce ground-state properties even in light nuclei with appropriate center-of-mass corrections, where it was described that the failure to accurately describe light nuclei (such as the deformations and shell evolution of helium and beryllium isotopic chains) has been attributed to limitations in the mean-field approximation itself\cite{Rong2023}.

Even for light nuclei, 
%{\color{red}\sout{many} 
all
%} 
parameter sets 
%{\color{red}\sout{(NS1, NS2, NS3, NL1, NL2, NL3, NL3*, NLB, NLC, and NLSH)}} 
remain stable within 0.3 MeV/A over 100 fm/c in second-order Runge Kutta, which is a duration longer than that used in fragmentation simulations. Among them, the parameter sets in which the binding energy per nucleon of the nuclear system, constructed immediately after nucleon sampling from the Wood-Saxon distribution, is well reproduced (Fig.~\ref{fig:repeat}) were NS2, NL3, NL3*, NLC, and NLSH.
It is difficult to identify a common feature among these parameter sets. However, it may be suggested that relatively low incompressibility is required to construct a stable ground-state nucleus in the QMD model (see Table II in the Appendix).
%{\color{red}
The fact that more stable ground states of nuclei can
be produced with parameter sets that have smaller incompressibility is likely indicated by the
reason that the binding energy does not vary significantly with density fluctuations. Since nuclear
incompressibility is defined as the second derivative of energy with respect to density, parameter
sets with equations of state in which energy does not change drastically in response to density
fluctuations correspond to this condition. In QMD simulations, it is considered that smaller nuclear
incompressibility was advantageous for numerical calculations that maintain the stability of nuclei
over a certain time step width.
%}

In the present study, mass dependence in the Gaussian wave packet width is introduced, as proposed in Ref.~\cite{Wang02}. Thus, the RMF model developed here may have greater generalizability for heavy nuclear collisions compared to the LIQMD model, which uses a fixed $L$ (1.26 fm$^2$). 
%{\color{red}
On the other hand, the use of the adaptive Gaussian wave packet results in different interaction ranges for different nuclei.
This poses a problem during the collision stage when the projectile and target differ. In this study, we used the weighted mean of 
$L$ between the projectile and the target, where the values of 
$L$ are quite limited—ranging from 1.072 (carbon projectile on hydrogen target) to 1.395 (carbon projectile on copper target).
It turns out that using a fixed value of 
$L$ may result in comparable fragmentation cross sections, making it suitable for hadron therapy applications.
%}
For further development, a density-dependent $L$, similar to the extended QMD approach \cite{Maruyama, Shi2024}, may be effective in describing ground-state properties across the entire nuclear mass range.

Fragmentation is well described by RQMD.RMF using the NS2 parameter set. In hadron therapy, uncertainties in dose distribution primarily arise from fragmentation in light elements that constitute the human body (H, C, N, O, P, and Ca together make up more than 90\% of the human body). Additionally, prompt gamma-ray production and gamma pair production, sequentially occurring due to fragmentation, provide information about the projectile range. Thus, nuclear fragmentation plays a significant role in hadron therapy. The LIQMD model has improved the distribution of positron-emitting radionuclides observed in PET analysis, presumably due to its enhanced forward fragment production\cite{Sato2024}. This study shows that the RQMD.RMF model is comparable to the LIQMD model in reproducing the fragmentation of 290 and 400 MeV/u carbon ion projectiles. Furthermore, the RQMD.RMF model describes fragmentation at lower energies (50 and 95 MeV/u) better than the LIQMD model. This suggests that the RQMD.RMF model is useful for hadron therapy analysis.

Other fragmentation models, including the Binary Cascade (BIC) model and the Liège Intranuclear Cascade (INCL) model, are also evaluated in this study (see Fig.~\ref{fig:MAEall} in Appendix), and among them, the RQMD.RMF model with NS2 is the most accurate within an energy range of 50–400 MeV/u for carbon ion beams interacting with light nuclei ($A \le 48$ in 50, and 95 MeV/u, and $A \le 63$ in 290, and 400 MeV/u). However, it should be noted that this result is not solely due to the use of the RMF model but also to QMD parameter adjustment, such as $L$ in QMD simulations. 

Recently developed QMD models, including LIQMD, use Eq.(\ref{RQMDE}), which assumes special relativity in the Hamiltonian, even when applying ``non-relativistic" nuclear models such as Skyrme parameter sets. The RQMD model has successfully described collective flows in the intermediate energy range. Therefore, it can be understood that the RMF model, which incorporates not only special relativity but also a covariant form in interactions via meson fields, extends naturally into QMD as a model that integrates nuclear matter properties more consistently.
%{\color{red}
The momentum-dependent interaction may play a more significant role at higher energies than those considered in the present study.
Therefore, further investigations are required to develop a more general model applicable to energy ranges where relativistic effects become increasingly important.
In addition, alternative parameter sets--such as models incorporating self-energy coupling for the 
$\omega$ meson \cite{Sugahara1994} or density-dependent couplings \cite{long2004}--should be explored to achieve a more accurate description of nucleus–nucleus collisions.
%}
%The density-dependent interaction, which was not included in this study, may be an important factor. Further investigations are expected to explore its effects in greater detail.

\section{\label{sec:conclusion}Conclusion}
In this study, the RQMD.RMF model, a quantum molecular dynamics approach based on relativistic mean field (RMF) theory, was evaluated for its accuracy in describing nuclear fragmentation processes within the energy range of 50–400 MeV/u, relevant to hadron therapy applications. A systematic assessment of 
%{\color{red}\sout{16} 
10
%} 
RMF parameter sets was conducted by analyzing ground-state properties, such as the mean squared radius, stability, and binding energy, as reproduced in QMD simulations. From this assessment, it was observed that the simultaneous reproduction of binding energy and nuclear size may require lower incompressibility as a nuclear matter property.

The NS2 parameter set was identified as the most suitable for describing stable nuclei across a wide mass range due to its compatibility with an adaptive Gaussian wave packet width. Then, the RQMD.RMF model using the NS2 parameter set was applied to simulate fragmentation cross-sections of carbon ion projectiles on various light nucleus targets--H, C, O, Al, and Ti at 50 and 95 MeV/u, and H, C, Al, and Cu at 290 and 400 MeV/u.

The simulation results demonstrated that the RQMD.RMF model is comparable to or outperforms other fragmentation models, such as the LIQMD model. These findings underscore the applicability of the RQMD.RMF model not only for accurately simulating nuclear fragmentation processes but also for practical applications in the analysis and planning of hadron therapy. By bridging the gap in understanding light nucleus collisions at intermediate energies, this study establishes the RQMD.RMF model as a valuable framework for advancing the precision and effectiveness of hadron therapy.
This model is expected to be implemented in the next future release of Geant4 version.

\begin{acknowledgments}
%{\color{red}
A.H. appreciates Dr. Niita for fruitful discussions regarding energy fluctuations and the adaptive Gaussian wave packet width.
%}
This work was supported by JSPS KAKENHI, Japan, Grant No. 23K07084.
\end{acknowledgments}

\appendix

%{\color{red}
\section{Derivatives of scalar, vector potentials, and self-energy terms}
The equation of motions shown in Eqs.~(\ref{EoM1}) and (\ref{EoM2}) require the derivatives of
$\Sigma_{s,k}, \Sigma_{v,k}^\nu, ({\rm SE})_k, f_k$, and $\beta_{k\nu}$ in $\vec{p}_i$ and $\vec{q}_i$. 
In this context, 
the useful formula can be derived as,
\onecolumngrid
\begin{align}
\sum_{k=1}^N {\Delta f_k} \frac{\partial \Sigma_{s,k}}{\partial \vec{p}_i} &= 
\sum_{k\ne i}^N 
\left[
{\Delta f_k}
D_{s,k}
\left(
\frac{\partial f_i}{\partial \vec{p}_i}\rho_{ki} + f_i\frac{\partial \rho_{ki} }{\partial \vec{p}_i}
\right)
+
{\Delta f_i}
D_{s,i}
%\frac{g_\sigma^2}{m_\sigma^2+2g_2\sigma_i+3g_3\sigma_i^2}
f_k\frac{\partial \rho_{ik} }{\partial \vec{p}_i}
\right],
\nonumber\\
\sum_{k=1}^N \frac{\partial \Sigma_{v,k}^0}{\partial \vec{p}_i} &= 
\sum_{k\ne i}^N
\frac{g_\omega^2}{m_\omega^2}
\left(
\frac{\partial \rho_{ki}}{\partial \vec{p}_i} + \frac{\partial \rho_{ik} }{\partial \vec{p}_i}
\right)
+
\sum_{k\ne i}^N
\frac{g_\rho^2}{m_\rho^2}
\left(
\frac{\partial \rho_{ki}}{\partial \vec{p}_i} + \frac{\partial \rho_{ik} }{\partial \vec{p}_i}
\right)\tau_{3,k}\tau_{3,i} \nonumber \\
& +
\sum_{k\ne i}^N
%\frac{1}{2}
\alpha \hbar c
\left(
\frac{1}{\sqrt{4\pi L}}e^{-\frac{\tilde{\vec{R}}_{ik}^2}{4L}}
-\frac{1}{2|\tilde{\vec{R}}_{ik}|}{\rm erf}({\frac{|\tilde{\vec{R}}_{ik}|}{\sqrt{4L}})}
\right)
\frac{\gamma_{ik}}{\tilde{\vec{R}}_{ik}^2}
\frac{\partial \tilde{\vec{R}}_{ik}^2}{\partial \vec{p}_i} Q_iQ_k
\nonumber\\
& +
\sum_{k\ne i}^N
%\frac{1}{2}
\alpha \hbar c
\frac{1}{|\tilde{\vec{R}}_{ik}|}
{\rm erf}({\frac{|\tilde{\vec{R}}_{ik}|}{\sqrt{4L}})}
\frac{\partial \gamma_{ik}}{\partial \vec{p}_i} Q_iQ_k
%\frac{\gamma_{ik}^2}{\sqrt{\vec{p}_i^2+m_i^2}+\sqrt{\vec{p}_k^2+m_k^2}}
%(\vec{\beta}_{ik}-\vec{\beta}_{ik}^2\vec{\beta}_i)
,
\nonumber\\
\sum_{k=1}^N {\Delta \vec{\beta}_k} \cdot \frac{\partial \vec{\Sigma}_{v,k}}{\partial \vec{p}_i} &= 
\sum_{k\ne i}^N
\frac{g_\omega^2}{m_\omega^2}
\left[
\frac{1}{\sqrt{\vec{p}_i^2+m_i^2}}
\left(
{\Delta \vec{\beta}_k}-
({\Delta \vec{\beta}_k}\cdot \vec{\beta}_i)\vec{\beta}_i
\right)\rho_{ki}
+
\left(
{\Delta \vec{\beta}_k}\cdot \vec{\beta}_i\frac{\partial \rho_{ki}}{\partial \vec{p}_i} 
+ {\Delta \vec{\beta}_i}\cdot \vec{\beta}_k\frac{\partial \rho_{ik} }{\partial \vec{p}_i}
\right)
\right] \nonumber \\
& +
\sum_{k\ne i}^N
\frac{g_\rho^2}{m_\rho^2}
\left[
\frac{1}{\sqrt{\vec{p}_i^2+m_i^2}}
\left(
{\Delta \vec{\beta}_k}-
({\Delta \vec{\beta}_k}\cdot \vec{\beta}_i)\vec{\beta}_i
\right)\rho_{ki}
+
\left(
{\Delta \vec{\beta}_k}\cdot \vec{\beta}_i\frac{\partial \rho_{ki}}{\partial \vec{p}_i} 
+ {\Delta \vec{\beta}_i}\cdot \vec{\beta}_k\frac{\partial \rho_{ik} }{\partial \vec{p}_i}
\right)
\right]\tau_{3,k}\tau_{3,i},
\nonumber\\
\sum_{k=1}^N\frac{\partial ({\rm SE})_k}{\partial \vec{p}_i} &= 
\frac{-\frac{1}{6}g_2-\frac{1}{4}g_3\sigma_i}{A_i} g_\sigma \sigma_i^2 \frac{\partial f_i}{\partial \vec{p}_i}
+
\sum_{k\ne i}^N 
\left(
G_k D_{s,k} \frac{\partial f_i}{\partial \vec{p}_i} \rho_{ki}
+G_k D_{s,k} f_i \frac{\partial \rho_{ki}}{\partial \vec{p}_i} 
+G_i D_{s,i} f_k \frac{\partial \rho_{ik}}{\partial \vec{p}_i} 
\right)
, \nonumber\\
\frac{\partial f_k}{\partial \vec{p}_i} &= - \frac{f_i}{\sqrt{\vec{p}_i^2+m_i^2}}\vec{\beta}_i\delta_{ki}, \nonumber\\
\frac{\partial \vec{\beta}_k}{\partial \vec{p}_i} &= 
\frac{1}{\sqrt{\vec{p}_i^2+m_i^2}}
\left(
I-
\vec{\beta}_i \vec{\beta}_i
\right) \delta_{ki}, \nonumber
%\sum_{k=1}^N \vec{\Sigma}_{v,k} \cdot \frac{\partial \vec{\beta}_k}{\partial \vec{p}_i} &= \nonumber
\end{align}
\noindent
where 
${\Delta f_k}$ and 
${\Delta \vec{\beta}_k}$ are ${\frac{1}{2}f_k-f_k^*}$ and ${\frac{1}{2}\vec{\beta}_k-\vec{\beta}_k^*}$, respectively,
and,
\begin{align}
D_{s,k} &= \frac{g_\sigma^2}{m_\sigma^2+2g_2\sigma_k+3g_3\sigma_k^2}, \nonumber\\
%G_{k} &= \frac{-\frac{1}{3}g_2-\frac{3}{4}g_3\sigma_k}{A_k} g_\sigma \sigma_k f_k
%-
%\frac{-\frac{1}{6}g_2-\frac{1}{4}g_3\sigma_k}{A_k^2} g_\sigma \sigma_k^2 f_k (g_2 + 2g_3\sigma_k)
G_{k} &= 
\frac{g_\sigma \sigma_k f_k}{A_k}
\left(
{-\frac{1}{3}g_2-\frac{3}{4}g_3\sigma_k}
%& \hspace{10mm}-
\frac{-\frac{1}{6}g_2-\frac{1}{4}g_3\sigma_k}{A_k} \sigma_k (g_2 + 2g_3\sigma_k)
\right),
\nonumber\\
A_{k} &= m_\sigma^2+g_2\sigma_k+g_3\sigma_k^2, \nonumber
\end{align}
are used. The derivatives in $\vec{q}_i$ can be derived as well.
%}
\twocolumngrid

\section{RMF parameter sets and the nuclear matter properties}
Table I indicates parameters of each RMF parameter set used in this study. 
Note that NS1$\sim$NS3 
%and MD1$\sim$MD4 given in Ref.\cite{Nara2019,Nara2020}
do not include $\rho$-meson, and therefore, in this study, we add it with the mass $m_{\rho} = 763$ MeV and the coupling constant $g_{\rho} = 7.5$. 
Corresponding nuclear matter properties including 
saturation density, binding energy, effective mass, incompressibility, and asymmetry energy are shown in Table II. 
%\subsection{\label{app:subsec}A: RMF parameter sets and the nuclear matter properties}

\begin{table*}[h]
\caption{\label{tab:table1} Parameter sets of relativistic mean field model used in this study.
$\dagger$ {\rm Unit\hspace{1mm}in\hspace{1mm}MeV.}}
\begin{ruledtabular}
\begin{tabular}{ccccccccccccccccc}
 % &\multicolumn{2}{c}{$D_{4h}^1$}&\multicolumn{2}{c}{$D_{4h}^5$}\\
  & {\scriptsize NS1\cite{Nara2019}} & {\scriptsize NS2\cite{Nara2019}} & {\scriptsize NS3\cite{Nara2019}} & {\scriptsize NL1\cite{Reinhard1986}} & {\scriptsize NL2\cite{Patra1993}} & {\scriptsize NL3\cite{Lalazissis1997}} & {\scriptsize NL3$^*$\cite{Lalazissis2009}} & {\scriptsize NLB\cite{Serot1992}} & {\scriptsize NLC\cite{Serot1992}} & {\scriptsize NLSH\cite{Sharma1993}} 
  %& \scriptsize MD1\cite{Nara2020} & \scriptsize MD2\cite{Nara2020} & \scriptsize MD3\cite{Nara2020} & \scriptsize MD4\cite{Nara2020} & \scriptsize TM1\cite{Sugahara1994} & \scriptsize PK1\cite{long2004}
  \\ 
  \hline
$g_\sigma$ &
6.448 &
7.902 &
8.864 &
10.14 &
9.111 &
10.22 &
10.09 &
9.696 &
9.750 &
10.44 
%&
%9.030 &
%9.233 &
%5.439 &
%4.059 &
%10.03 &
%10.32 
\\
$^\dagger g_2$ &
-76 &
88.62 &
4.382 &
24.34 &
4.608 &
20.86 &
21.62 &
4.055 &
25.34 &
13.82 
%&
%8.436 &
%8.024 &
%-31.18 &
%-320.6 &
%14.47 &
%16.34 
\\
$g_3$ &
2038 &
131.9 &
162.4 &
-217.6 &
82.70 &
-173.3 &
-180.9 &
10.0 &
-200.0 &
-95.0 
%&
%40.0 &
%33.12 &
%2347 &
%16104 &
%3.710 &
%-59.99 
\\
$^\dagger m_\sigma$ &
550.0 &
550.0 &
550.0 &
492.3 &
504.9 &
508.2 &
502.6 &
510.0 &
500.8 &
526.1 
%&
%550.0 &
%550.0 &
%550.0 &
%550.0 &
%511.2 &
%514.1 
\\
$g_\omega$ &
6.859 &
6.859 &
10.07 &
13.29 &
11.49 &
12.87 &
12.81 &
12.59 &
12.20 &
12.95 
%&
%6.740 &
%3.888 &
%0 &
%5.632 &
%12.61 &
%13.01 
\\
%$c_3$ &
%0 &
%0 &
%0 &
%0 &
%0 &
%0 &
%0 &
%0 &
%0 &
%0 &
%0 &
%0 &
%0 &
%0 &
%71.31 &
%55.64\\
$^\dagger m_\omega$ &
783.0 &
783.0 &
783.0 &
795.4 &
780.0 &
782.5 &
782.6 &
783.0 &
783.0 &
783.0 
%&
%783.0 &
%783.0 &
%783.0 &
%783.0 &
%783.0 &
%784.3 
\\
$g_\rho$ &
7.5 &
7.5 &
7.5 &
9.951 &
11.01 &
8.948 &
9.150 &
8.544 &
8.66 &
8.766 
%&
%7.5 &
%7.5 &
%7.5 &
%7.5 &
%9.264 &
%9.059 
\\
$^\dagger m_\rho$ &
763 &
763 &
763 &
763 &
763 &
763 &
763 &
770 &
770 &
763 
%&
%763 &
%763 &
%763 &
%763 &
%763 &
%763 
\\
%$\bar{g}_\sigma$ &
%0 &
%0 &
%0 &
%0 &
%0 &
%0 &
%0 &
%0 &
%0 &
%0 &
%3.186 &
%2.502 &
%7.711 &
%5.544 &
%0 &
%0 
%\\
%$\bar{g}_\omega$ &
%0 &
%0 &
%0 &
%0 &
%0 &
%0 &
%0 &
%0 &
%0 &
%0 &
%8.896 &
%10.43 &
%11.22 &
%3.926 &
%0 &
%0 \\
%$\bar{g}_\rho$ &
%0 &
%0 &
%0 &
%0 &
%0 &
%0 &
%0 &
%0 &
%0 &
%0 &
%0 &
%0 &
%0 &
%0 &
%0 &
%0 \\
%$^\dagger \Lambda_\sigma$ &
%- &
%- &
%- &
%- &
%- &
%- &
%- &
%- &
%- &
%- &
%641 &
%489.7 &
%1702 &
%704 &
%- &
%- \\
%$^\dagger \Lambda_\omega$ &
%- &
%- &
%- &
%- &
%- &
%- &
%- &
%- &
%- &
%- &
%1841 &
%2489 &
%1898 &
%4252 &
%- &
%- \\
%$^\dagger \Lambda_\rho$ &
%- &
%- &
%- &
%- &
%- &
%- &
%- &
%- &
%- &
%- &
%- &
%- &
%- &
%- &
%- &
%- \\
\end{tabular}
%\footnote{$\dagger$ {\rm Unit\hspace{1mm}in\hspace{1mm}MeV.}}
\end{ruledtabular}
%\end{table*}
%\begin{table*}[h]
\caption{\label{tab:table2} Properties in nuclear matter.
$\rho_0$: Saturation density [fm$^{-3}$], $E/A$: Binding energy [MeV], 
$m^*$: Effective mass, $K$: Incompressibility [MeV], $J$: Asymmetry energy [MeV].}
\begin{ruledtabular}
\begin{tabular}{ccccccccccccccccc}
 % &\multicolumn{2}{c}{$D_{4h}^1$}&\multicolumn{2}{c}{$D_{4h}^5$}\\
    & {\scriptsize NS1\cite{Nara2019}} & {\scriptsize NS2\cite{Nara2019}} & {\scriptsize NS3\cite{Nara2019}} & {\scriptsize NL1\cite{Reinhard1986}} & {\scriptsize NL2\cite{Patra1993}} & {\scriptsize NL3\cite{Lalazissis1997}} & {\scriptsize NL3$^*$\cite{Lalazissis2009}} & {\scriptsize NLB\cite{Serot1992}} & {\scriptsize NLC\cite{Serot1992}} & {\scriptsize NLSH\cite{Sharma1993}} %& {\scriptsize MD1\cite{Nara2020}} & {\scriptsize MD2\cite{Nara2020}} & {\scriptsize MD3\cite{Nara2020}} & {\scriptsize MD4\cite{Nara2020}} & {\scriptsize TM1\cite{Sugahara1994}} & {\scriptsize PK1\cite{long2004}}
    \\ \hline
$\rho_0$ &
0.168 &
0.168 &
0.168 &
0.152 &
0.146 &
0.148 &
0.150 &
0.148 &
0.149 &
0.146 
%&
%0.168 &
%0.168 &
%0.168 &
%0.168 &
%0.145 &
%0.148 
\\
$E/A$ &
-15.97 &
-15.97 &
-15.91 &
-16.43 &
-17.02 &
-16.24 &
-16.31 &
-15.77 &
-15.77 &
-16.36 
%&
%-15.77 &
%-15.86 &
%-15.70 &
%-15.99 &
%-16.26 &
%-16.27 
\\
$m^*$ &
0.83 &
0.83 &
0.70 &
0.57 &
0.67 &
0.60 &
0.59 &
0.61 &
0.63 &
0.60 
%&
%0.65 &
%0.65 &
%0.65 &
%0.82 &
%0.63 &
%0.61 
\\
$K$ &
380.3 &
209.5 &
379.0 &
211.5 &
399.6 &
270.8 &
257.6 &
420.6 &
223.9 &
355.6 
%&
%379.2 &
%380.7 &
%381.6 &
%210.5 &
%281.7 &
%281.9 
\\
$J$ &
30.07 &
30.01 &
32.34 &
43.49 &
45.13 &
37.36 &
38.64 &
35.0  &
35.33 &
36.13 
%&
%32.99 &
%33.13 &
%32.77 &
%30.07 &
%37.30 &
%37.60 
\\
\end{tabular}
\end{ruledtabular}
\end{table*}

\section{Comparison with experimental data}
%\subsection{\label{app:subsec}Details of angular distribution in 50 and 95 MeV/u}

\begin{figure*}[h]
\includegraphics[width=15cm]{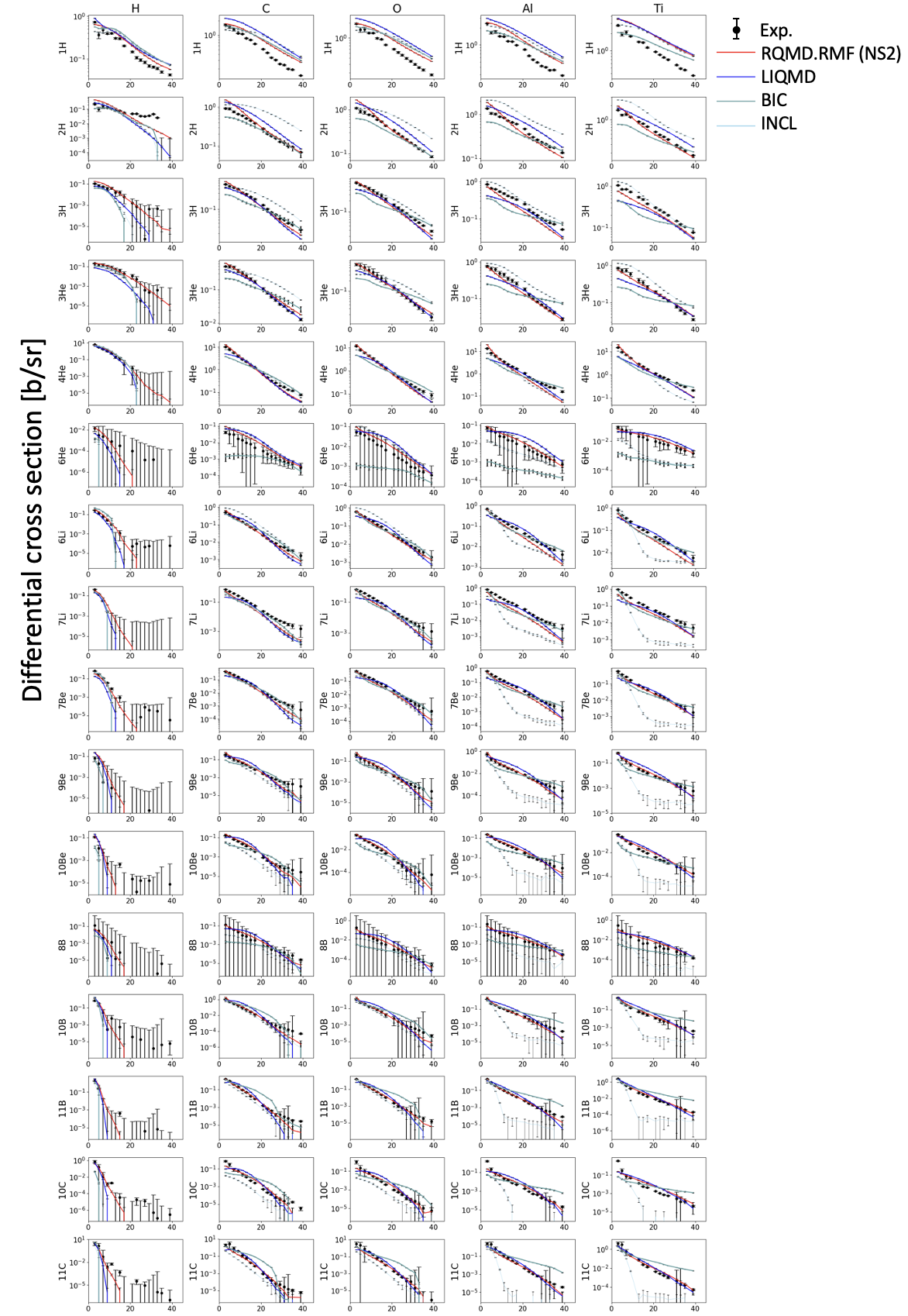}% Here is how to import EPS art
\caption{\label{fig:50Angle}Differential cross-section of carbon ion 50 MeV/u beams for H, C, O, Al, and Ti targets. The results of four fragmentation models, RQMD.RMF (NS2 parameter set), Light Ion QMD (LIQMD)\cite{Sato2022}, Binary Cascade (BIC)\cite{Folger2004}, and Liège Intranuclear Cascade (INCL)\cite{Boudard2013} models, are compared to the experiment\cite{Divay2017}.}
\end{figure*}

\begin{figure*}
\includegraphics[width=15cm]{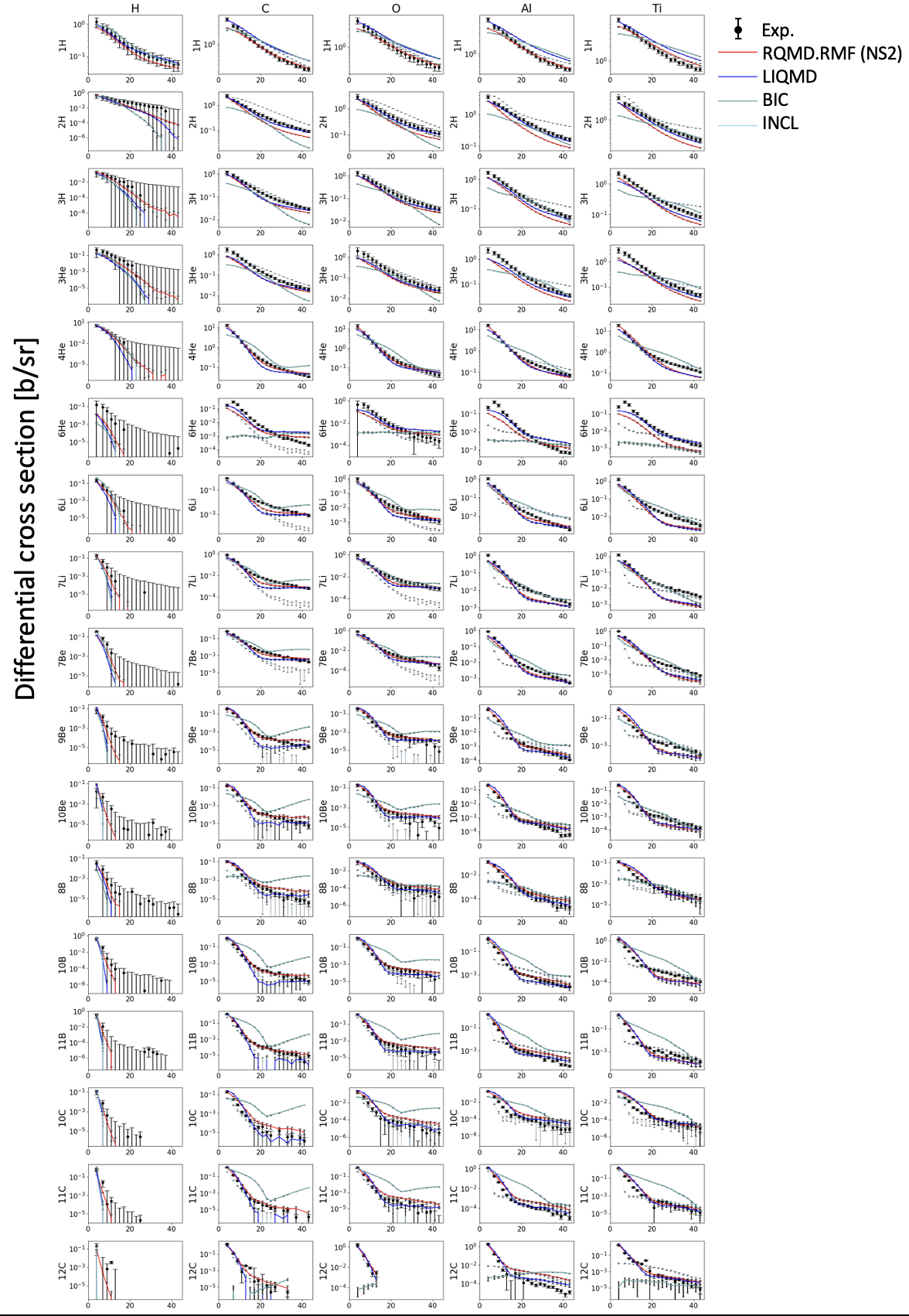}% Here is how to import EPS art
\caption{\label{fig:95Angle}
Differential cross-section of carbon ion  95 MeV/u beams for H, C, O, Al, and Ti targets. The results of four fragmentation models, RQMD.RMF (NS2 parameter set), Light Ion QMD (LIQMD)\cite{Sato2022}, Binary Cascade (BIC)\cite{Folger2004}, and Liège Intranuclear Cascade (INCL)\cite{Boudard2013} models, are compared to the experiment\cite{Dudouet2013, Dudouet2014}.}
\end{figure*}

Differential cross-sections of carbon ion beams with 50 and 95 MeV/u in each fragment for H, C, O, Al, and Ti targets are given in Figs.~\ref{fig:50Angle} and \ref{fig:95Angle}, respectively. 
The results of four fragmentation models, RQMD.RMF (NS2 parameter set), Light Ion QMD (LIQMD)\cite{Sato2022}, Binary Cascade (BIC)\cite{Folger2004}, and Liège Intranuclear Cascade (INCL)\cite{Boudard2013} models implemented in Geant4 ver.11.2, are simultaneously plotted for comparison.

\begin{figure*}
\includegraphics[width=15cm]{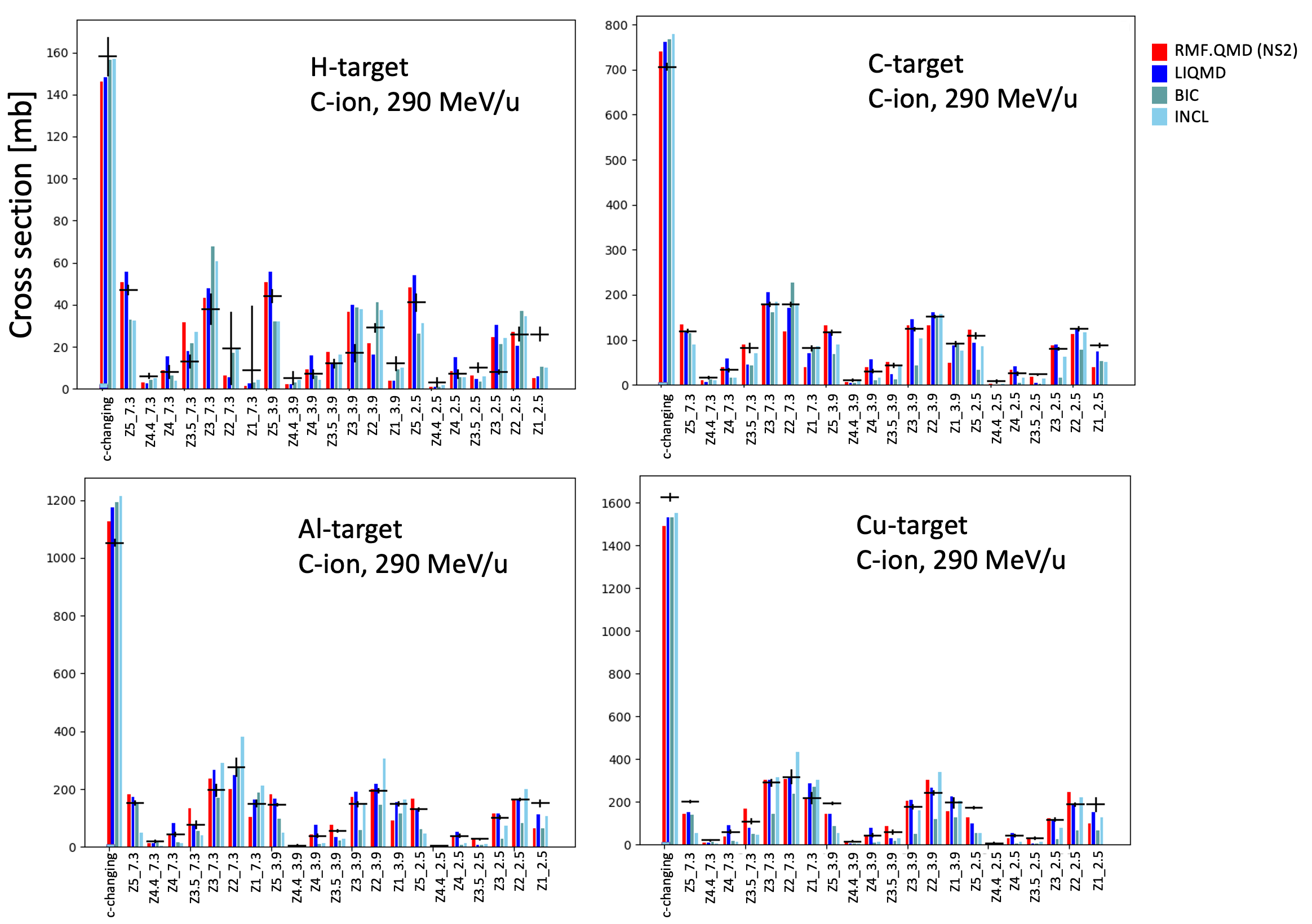}% Here is how to import EPS art
\caption{\label{fig:290each}
Cross-sections of carbon ion 290 MeV/u beams for H, C, Al, and Cu targets; c-changing means charge-changing cross section, Z(\#1)\_(\#2) means the production cross section of effective charge (\#1) with acceptance angle of (\#2). 
The results of five fragmentation models, RQMD.RMF (NS2 parameter set), Light Ion QMD (LIQMD)\cite{Sato2022}, Binary Cascade (BIC)\cite{Folger2004}, and Liège Intranuclear Cascade (INCL)\cite{Boudard2013} models are compared to the experiment\cite{Zeitlin2007}.
%The result of PHITS is taken from Ref.\cite{Zeitlin2007}.
}
\includegraphics[width=15cm]{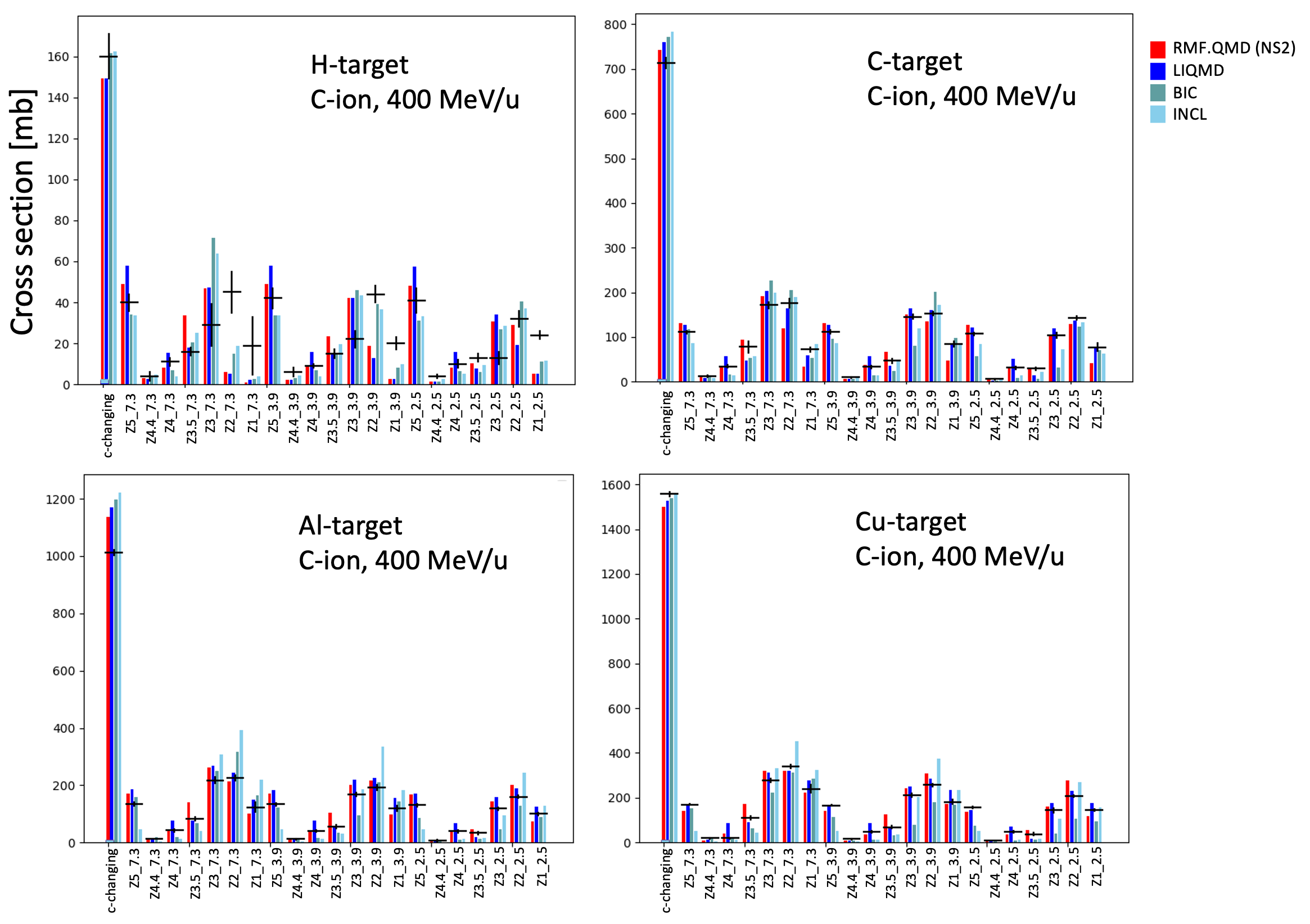}% Here is how to import EPS art
\caption{\label{fig:400each}
Cross-sections of carbon ion 400 MeV/u beams for H, C, Al, and Cu targets; c-changing means charge-changing cross section, Z(\#1)\_(\#2) means the production cross section of effective charge (\#1) with acceptance angle of (\#2). 
The results of five fragmentation models, RQMD.RMF (NS2 parameter set), Light Ion QMD (LIQMD)\cite{Sato2022}, Binary Cascade (BIC)\cite{Folger2004}, and Liège Intranuclear Cascade (INCL)\cite{Boudard2013} models  are compared with the experiment\cite{Zeitlin2007}.
%The result of PHITS is taken from Ref.\cite{Zeitlin2007}.
}
\end{figure*}

On the other hand, Figs.~\ref{fig:290each} and \ref{fig:400each} show the cross-sections of 290 MeV/u and 400 MeV/u carbon ion beams, respectively, for H, C, Al, and Cu targets, where c-changing means charge-changing cross section, Z(\#1)\_(\#2) means the production cross section of effective charge (\#1) with acceptance angle of (\#2). 
The results of five fragmentation models, RQMD.RMF (NS2), LIQMD, BIC, INCL models implemented in Geant4 ver.11.2 are compared to the experiment.
%It should be noted that the result in PHITS is not that of the latest model, but is bought from Ref.\cite{Zeitlin2007}.

\begin{figure*}
\includegraphics[width=15cm]{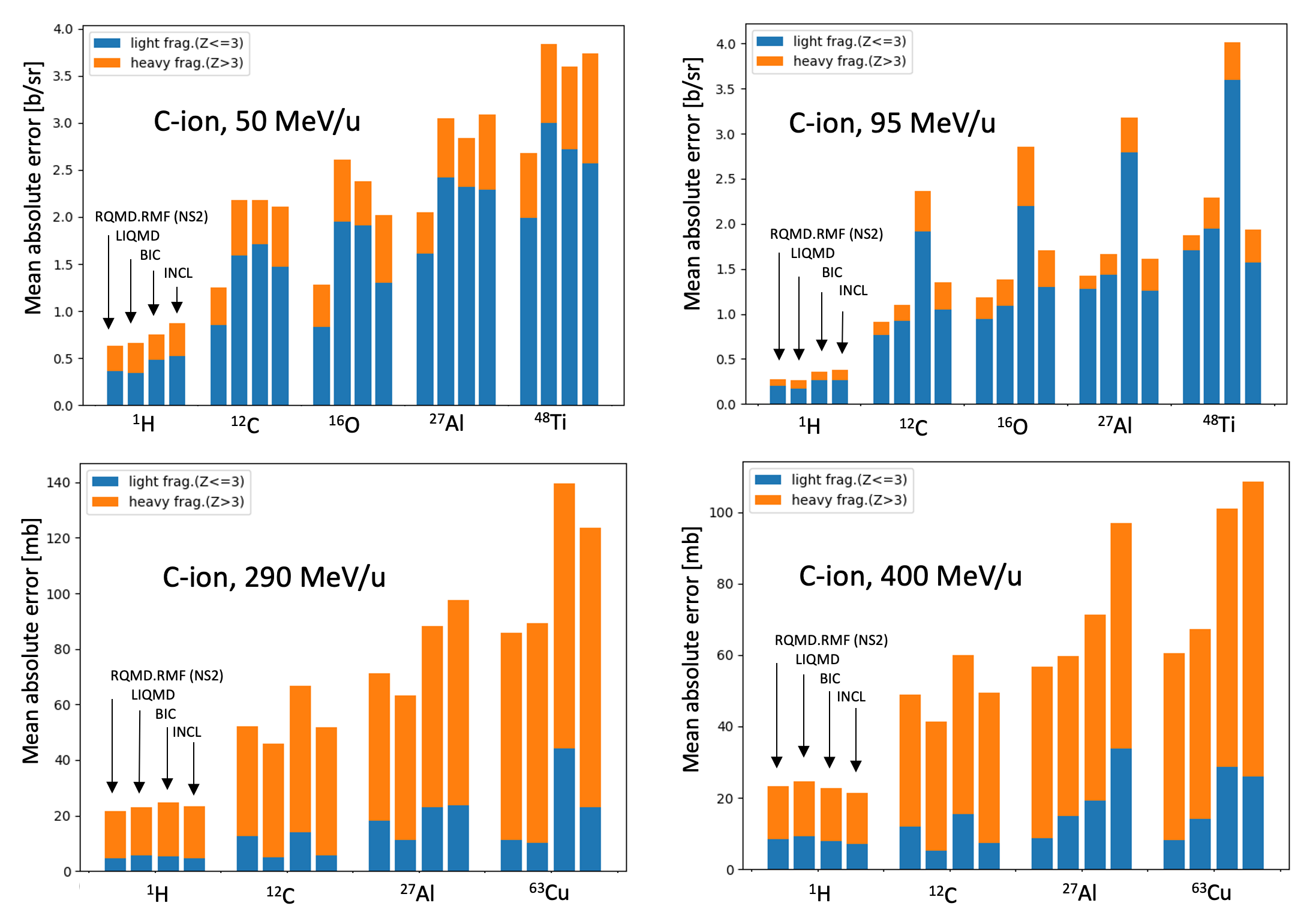}% Here is how to import EPS art
\caption{\label{fig:MAEall}
Mean absolute errors (MAEs) obtained from Figs.~\ref{fig:50Angle}$\sim$\ref{fig:400each}.
Upper left: 50 MeV/u, 
upper right: 95 MeV/u,
lower left: 290 MeV/u,
lower right: 400 MeV/u.
The figures shows the MAE results of RQMD.RMF (NS2 parameter set), Light Ion QMD (LIQMD)\cite{Sato2022}, Binary Cascade (BIC)\cite{Folger2004}, and Liège Intranuclear Cascade (INCL)\cite{Boudard2013} models
%from the left, whereas The upper figures shows the MAE result of RQMD.RMF (NS2 parameter set), LIQMD, BIC, INCL, and PHITS models
from the left.
}
\end{figure*}

The MAEs in each fragmentation model are compared in Fig.\ref{fig:MAEall}, which are obtained from the result shown in Figs.\ref{fig:50Angle}$\sim$\ref{fig:400each}.

%\subsection{\label{app:subsec}Details of mean absolute error in 290 and 400 MeV/u}

%\nocite{*}

\bibliography{apssamp}% Produces the bibliography via BibTeX.

\end{document}